\documentclass[structabstract]{aa}
\pdfoutput=1
\usepackage{graphicx}
\usepackage{subfigure}
\usepackage{rotating}
\usepackage{float}
\usepackage{enumitem}
\usepackage{har2nat}
\usepackage{amsmath}
\usepackage{txfonts}
\usepackage{xcolor}

\newcommand{\Msun}{M$_{\odot}$}

\newcommand{\SI}{[\ion{S}{II}]$\lambda$6731}
\newcommand{\Ox}{[\ion{O}{I}]$\lambda$6300}

\newcommand{\Ha}{H$\alpha$}
\newcommand{\SII}{[\ion{S}{II}]$\lambda\lambda$6716,6731}
\newcommand{\NII}{[\ion{N}{II}]$\lambda$6583}

\newcommand{\CaII}{[\ion{Ca}{II}]$\lambda$7291}

\newcommand{\km}{km~s$^{-1}$}

\begin{document}

\title{The Morphology of the HD~163296 jet as a window on its planetary system
\thanks{Based on observations collected with MUSE at the Very Large Telescope on Cerro Paranal (Chile),
operated by the European Southern Observatory (ESO). Program ID: 099.C-0214(A).}}

\author{A. Kirwan \inst{1}
  \and 
  A. Murphy. \inst{1}
  \and
  P.C. Schneider \inst{2}
  \and
 E.T. Whelan \inst{1}
  \and 
  C. Dougados \inst{3} 
  \and
  J. Eisl{\"o}ffel \inst{4}} 

\institute{Maynooth University Department of Experimental Physics, National University of Ireland Maynooth, Maynooth Co. Kildare, Ireland
 \and 
Hamburger Sternwarte, Universit{\"a}t Hamburg, Gojenbergsweg 112, 21029 Hamburg, Germany
\and
Univ. Grenoble Alpes, CNRS, IPAG, 38000 Grenoble, France
\and
Th{\"u}ringer Landessternwarte Tautenburg, Sternwarte 5, D-07778 Tautenburg, Germany}
\titlerunning{The morphology of the HD~163296 jet as a window on its planetary system} 
\date{}

\abstract {HD~163296 is a Herbig Ae star which drives a bipolar knotty jet with a total length of $\sim$6000~au. Strong evidence exists that the disk of HD~163296 harbors planets. {Studies have shown that the presence of companions around jet-driving stars could affect the morphology of the jets. This includes a `wiggling' of the jet axis and a periodicity in the positions of the jet knots.}}{In this study we investigate the morphology (including the jet width and axis position) and proper motions of the HD~163296 jets, and use our results to better understand the whole system.}{This study is based on optical integral-field spectroscopy observations obtained with VLT/MUSE in 2017. Using spectro-images and position velocity diagrams extracted from the MUSE data cube, we investigated the number and positions of the jet knots. A comparison was made to X-Shooter data collected in 2012 and the knot proper motions were estimated. The jet width and jet axis position with distance from the star were studied from the extracted spectro-images. {This was done using Gaussian fitting to spatial profiles of the jet emission extracted perpendicular to the position angle of the jet. The centroid of the fit is taken as the position of the jet axis.}}{We observe the merging of knots and identify two previously undetected knots.  We find proper motions that are broadly in agreement with previous studies. The jet width increases with distance from the source and we measure an opening angle of $\sim$ 5$^{\circ}$ and $2.5^{\circ}$ for the red and blue lobes, respectively. {Measurements of the jet axis position, derived from Gaussian centroids of transverse intensity profiles, reveal a similar pattern of deviation in all forbidden emission lines along the first 20 arc seconds of the jets. This result is interpreted as being due to asymmetric shocks and not due to a wiggling of the jet axis.}}{The number of new knots detected and their positions challenge the 16-year knot ejection periodicity proposed in prior studies, arguing for a more complicated jet system than was previously assumed. {We use the non-detection of a jet axis wiggling} to rule out companions with a mass $>$0.1~\Msun\ and orbits between 1~au and 35~au. Any object inferred at these distances using other methods must be a brown dwarf or planet, otherwise it would have impacted the jet axis position. Both the precession and orbital motion scenarios are considered. Overall it is concluded that it is difficult to detect planets with orbits $>$1~au through a study of the jet axis.}


\maketitle

\section{Introduction} 

Jets from young stellar objects (YSOs) are collimated outflows of matter launched from the system in the early stages of the formation process. Their formation is theorised to be related to magnetohydrodynamic (MHD) processes in the rotating star-disk system and thus they are strongly connected to the accretion disk in which planets may form \citep{RayNewAR, Whelan2014}. Stellar jets are typically made up of a string of shock-heated nebulosities or knots prominently observed in forbidden emission lines (FELs), each one likely corresponding to a different ejection event \citep{Whelan2012}. Thus, these knots trace the mass loss history of the star~\citep{ellerbroek2014, Whelan2014}. As well as offering insight into the evolution of the YSO through its mass loss, the morphology of the jet can also open a window into the unresolved central engine of the system \citep{Murphy2021}. In particular, a change in the position of the jet axis with time can indicate an orbital motion of the outflow source or precession of the jet axis due to an inner disk warp, for example~\citep{Crespo2020, Lai}. Thus, a study of the evolution of YSO jet axes can point to undetected stellar or brown dwarf companions or possibly even planets \citep{Murphy2021}. Periodic ejection events could also be caused by an interaction between the driving source and a stellar or planetary companion \citep{ellerbroek2014, Muzerolle2013}. Spectro-imaging is a powerful tool to study and discover protostellar jets~\citep{Schneider2020aap}; is it a particularly useful tool for studying YSO jet axes and has frequently provided observations of jet wiggling~\citep{Dougados2000, Whelan2010, Murphy2021, Erkal2021AA}. 

\begin{figure*}
\centering
\includegraphics[width=15.0cm, trim={1cm 0cm 1cm 0cm}, clip=true]{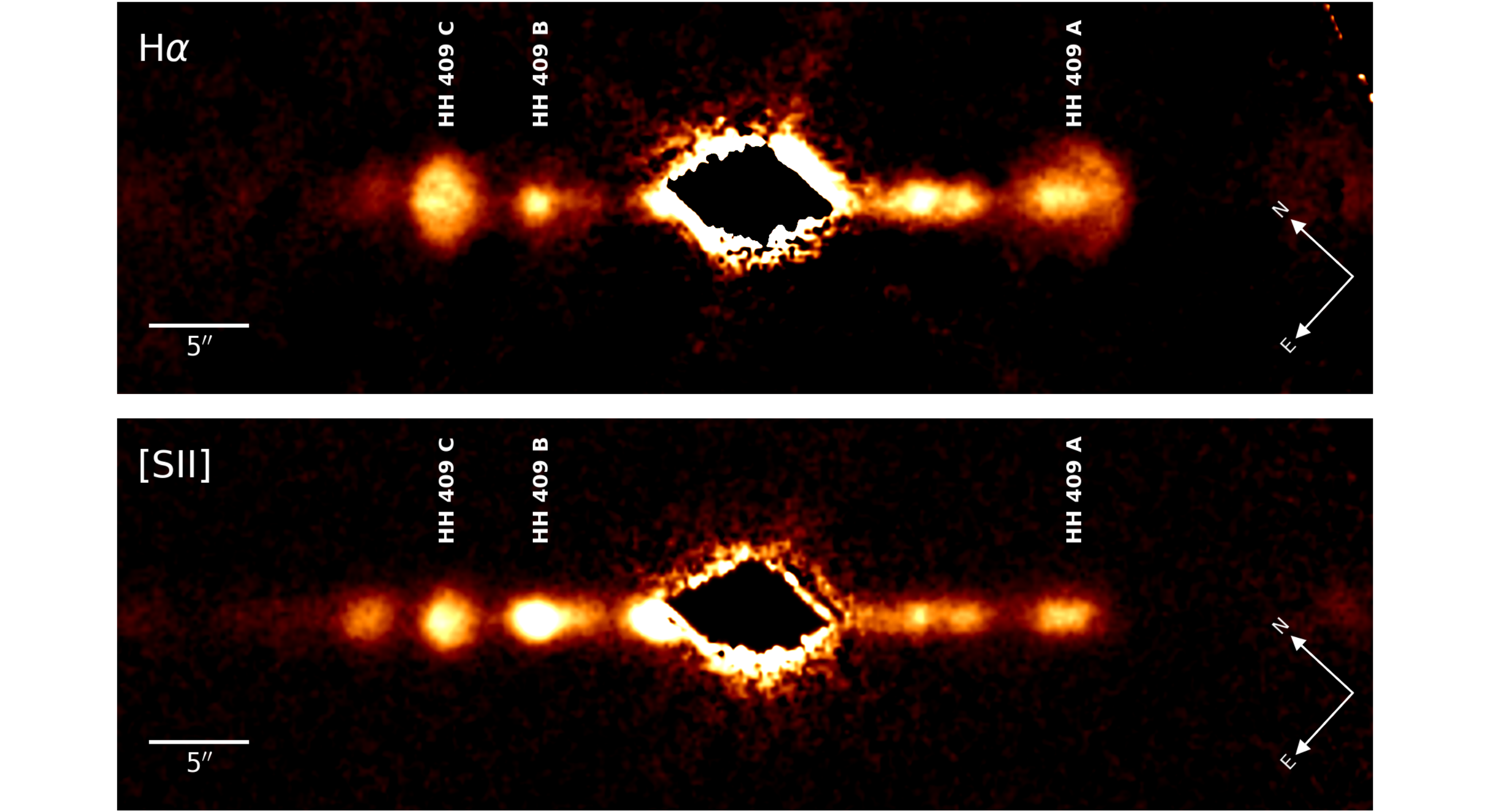}
     \caption{Images of the jet in \Ha~and \SI, with the prominent HH objects labelled. The image corresponds to a $\sim$12.5\AA~wide spectral bin centred on the peak of the emissions. It is rotated by an angle of 47.3$^{\circ}$ with respect to the plane of the sky. We note the lack of emission in the first few arcseconds due to the saturation effects discussed in Section 2.}
  \label{spectro_image} 
\end{figure*}

The focus of this paper is a morphological study of the bipolar knotty jet from the YSO HD~163296. HD~163296 is a $\sim$1.9~M$_{\odot}$, $\sim$4.4~Myr Herbig Ae star located at a distance of 101.5 $\pm$ 1.2~pc \citep{flaherty2015, pinte2018}. It was first discovered to be driving a bipolar jet using the Hubble Space Telescope (HST)~\citep{grady2000, devine2000} and subsequent studies have put the length of the jet at $\sim$ 3000~au on either side of the source. The jet is made up of a series of at least ten knots with radial velocities of 100 - 300~km~s$^{-1}$~\citep{wassell2006, ellerbroek2014}. HD~163296 is surrounded by a large circumstellar disk ($d \sim$ 750~au) with an inclination angle of 41.4 $\pm$ 0.3$^{\circ}$ and position angle (PA) of 132.2 $\pm$ 0.3$^{\circ}$~\citep{Rich2019ApJ}. The observation of rings in the disk led to the conjecture that these rings could be carved out by planets~\cite{grady2000, isella2016}, although the observations of~\citet{grady2000} were later interpreted as scattered light by~\citet{wisniewski2008}. More recent studies have proposed the existence of two Jupiter-mass planets using kinematical models and emission line analysis~\citep{teague2018, pinte2018}. 

Here, we present a unique view of the HD~163296 jet obtained with the Multi Unit Spectroscopic Explorer (MUSE) on the European Southern Observatory's Very Large Telescope~\citep[VLT, see][]{Bacon2010}. The simultaneous spectral and optical imaging capabilities allow for examination of the morphological and kinematic structure of the jet in various velocity regimes and emission lines, and the large field of view (FOV) of 60\arcsec\, $\times$ 60\arcsec\, further permits such study on extended regions of the jet. Due to suggestions of planet formation in the disk of this YSO, a particular aim of this work is to explore the evolution of the jet axis for evidence of companions. In the text we use the term `jet axis wiggling' to refer to a real, physical change in the position of the jet axis, which could possibly be due to a companion or an inner disk warp. Thus the results presented here are particularly focused on a mapping of the jet width and jet axis position. In addition, we also report on changes in the morphology of the jet and on the proper motions of the knots. The observations, reduction, and analysis of the resultant data are discussed in Section~\ref{sec:reduction}. The results of the morphological analysis and proper motions are described in Section~\ref{sec:results}. In Section~\ref{sec:disconc} the evolution of the HD~163296 jet axis is further discussed and the conclusions given in Section~\ref{sec:conc}. 

\section{Observations, data reduction, and analysis}
\label{sec:reduction}

The MUSE Observations of HD~163296 were performed on October 2nd 2017 under Program ID 099.C-0214(A) (PI: C. Schneider). The instrument was operated in Wide-Field Mode (WFM) without adaptive optics correction and the seeing was $\sim$0\farcs8. The position angle (PA) of the IFU was rotated between the exposures to improve image sampling and we concentrate on the exposures with $t_{exp}=100\,$s, which comprise 75\% of the on-target time\footnote{A few  exposures with shorter integration times exist, but we focus on the extended jet here so that adding more but short exposures does not noticeably improve the data quality.}. The total on-source integration time for the long-exposure cubes was $1200\,$s.  The ESO MUSE pipeline (version 2.6; see \citealt{Weilbacher2020A&A} for discussion of the pipeline and reduction process) was used to reduce the data along with the standard calibration files. We performed the reduction according to the standard pipeline parameters except for the following issue. Due to the central star's optical brightness (V=6.8\,mag), the detector suffers from saturation effects in the immediate region around the star. Scattering within the detector adds excess noise in parts of the IFU and we discard these detector parts during the data reduction as the observing setup with multiple detector PAs results in full spatial coverage in the outer jet regions (see Figure~\ref{spectro_image}). The final cube was created by stacking the individual exposures into smaller cubes spanning fixed wavelength ranges, which were then combined into a single data cube covering the entire wavelength range of MUSE (4800-9300\AA) with an average spectral resolution of $\sim5300$ in our particular region of interest. Due to the rotation of the exposures in the reduction process, the final cube has an approximately circular sky coverage with a diameter of $\sim$84\arcsec.

\begin{figure*}
    \centering
    \includegraphics[width=0.6\textwidth, trim={0cm 0cm 0cm 0cm}, clip=true]{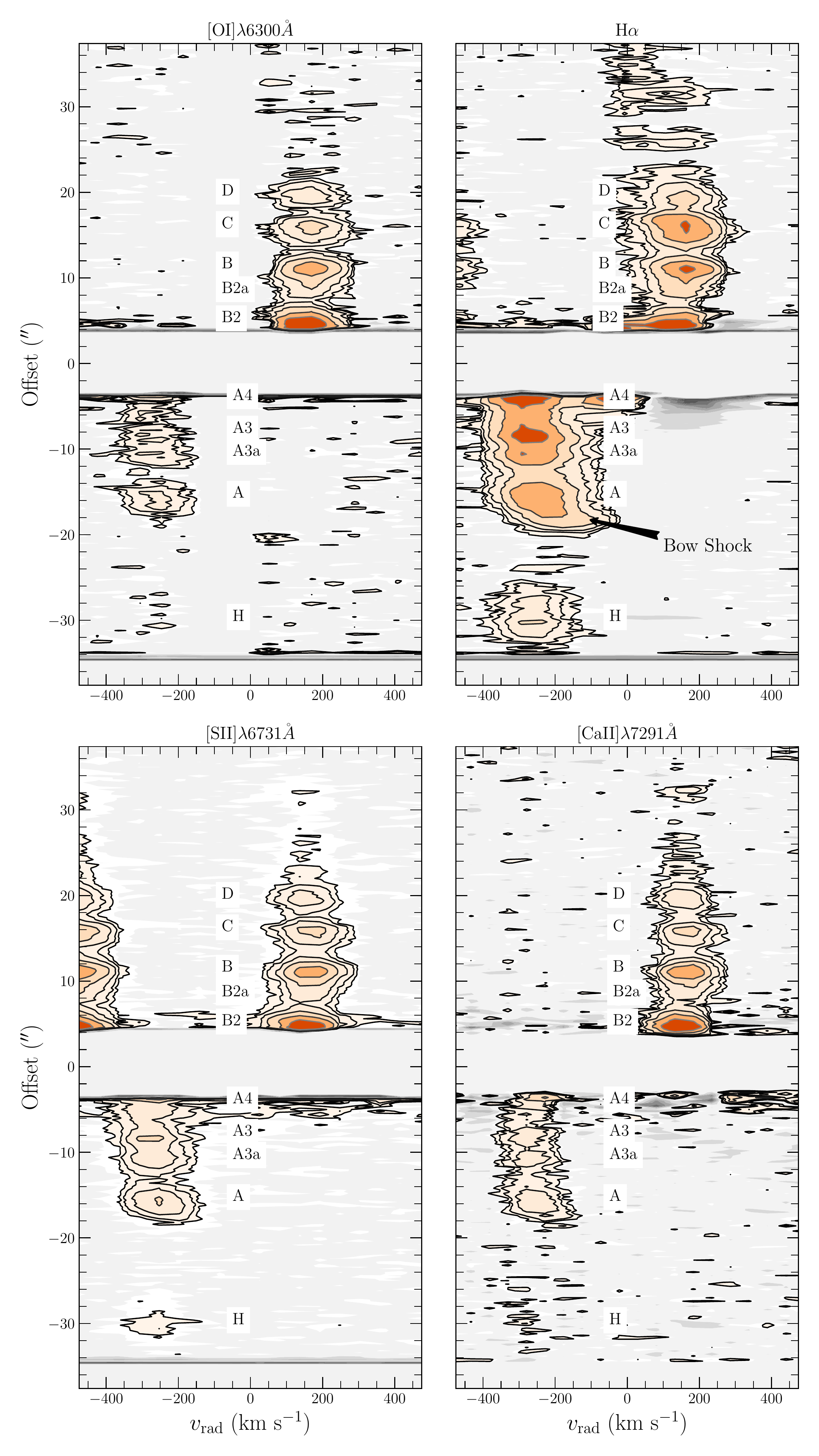}
    \caption{Position-velocity diagrams of the \Ox~(top left), \Ha~(top right), \SI~(bottom left), and \CaII~(bottom right) emissions of the HD 163296 outflow, corrected for the local standard of rest. The colour contours represent a 3$\sigma$ detection, and the background contours are on a log scale up to 2$\sigma$. The bow shock in knot A is visible in \Ha, \SI, and \CaII\, (see Table~\ref{knot_table}). The emission at the far edge of the \SI\, line is the red-shifted lobe of the [\ion{S}{ii}]$\lambda$6716 line.}
    \label{fig:pvs1}
\end{figure*}

\begin{figure*}
	\centering
\includegraphics[width=0.8\textwidth, trim={0cm 0.2cm 0cm 1cm}, clip=true]{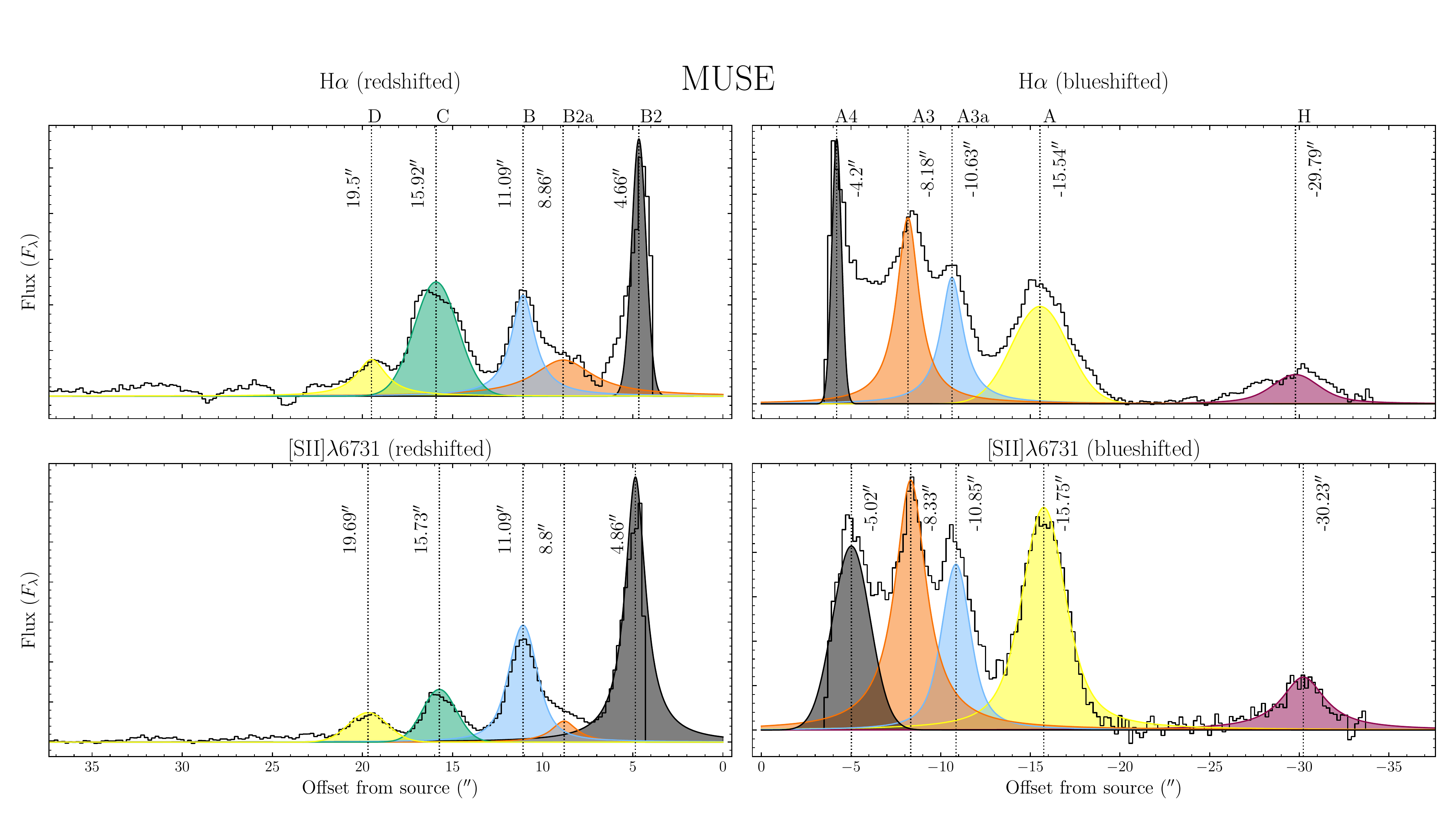}
\includegraphics[width=0.8\textwidth, trim={0cm 0.2cm 0cm 1cm}, clip=true]{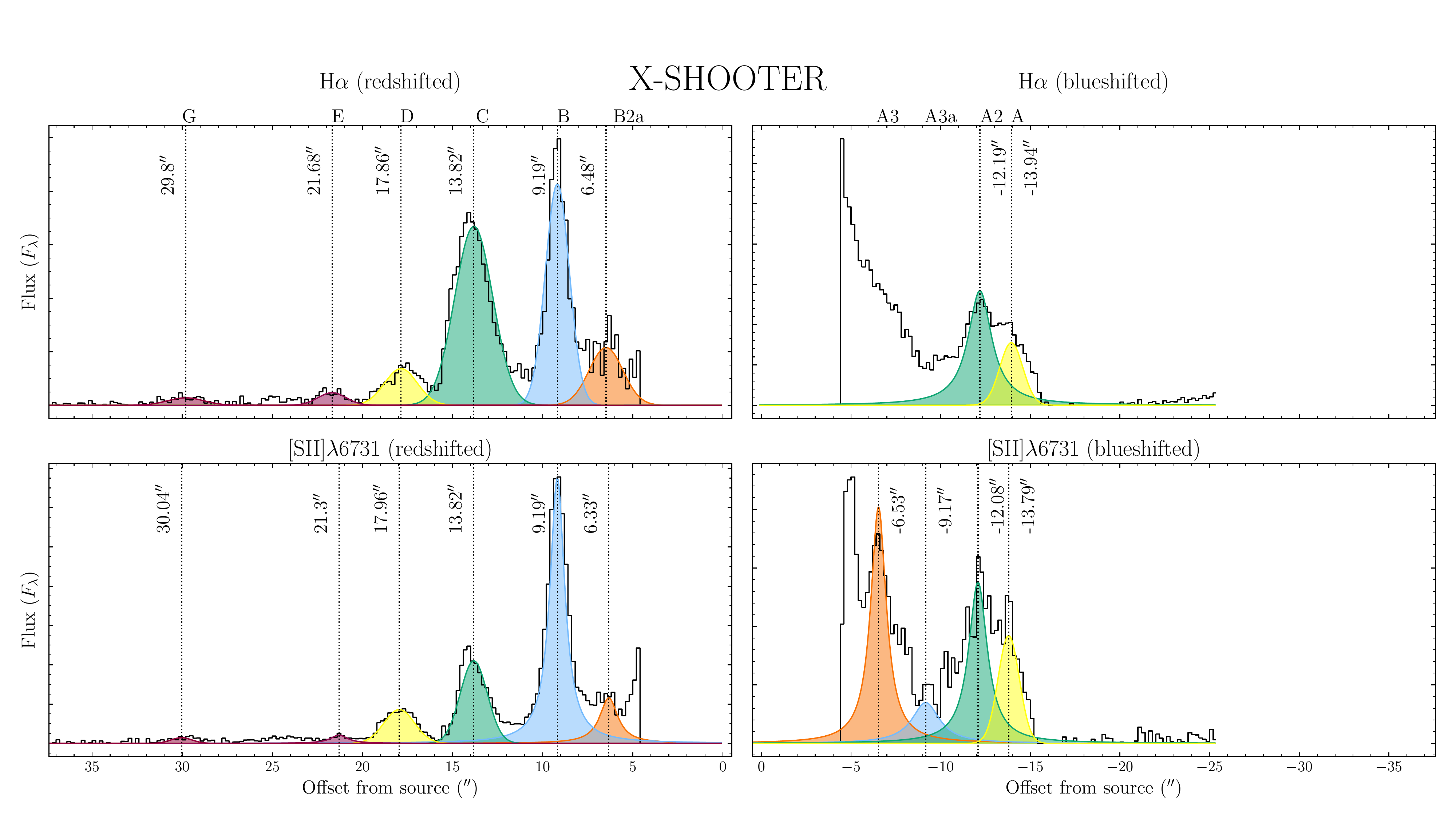}
    \caption{Spatial profiles of the jet in \Ha~and \SI~emission lines showing the a) MUSE data and b) X-Shooter data fitted component-wise. We note a potential unreported knot at $\sim9.1$\arcsec~in the blue \SI~line of the X-Shooter data which we believe to be seen as A3a in the MUSE data.}
    \label{fig:full_line_fits}
\end{figure*}

\begin{figure*}
	\centering
    \includegraphics[width=0.7\textwidth, trim={0cm 0.5cm 0cm 0cm}, clip=true]{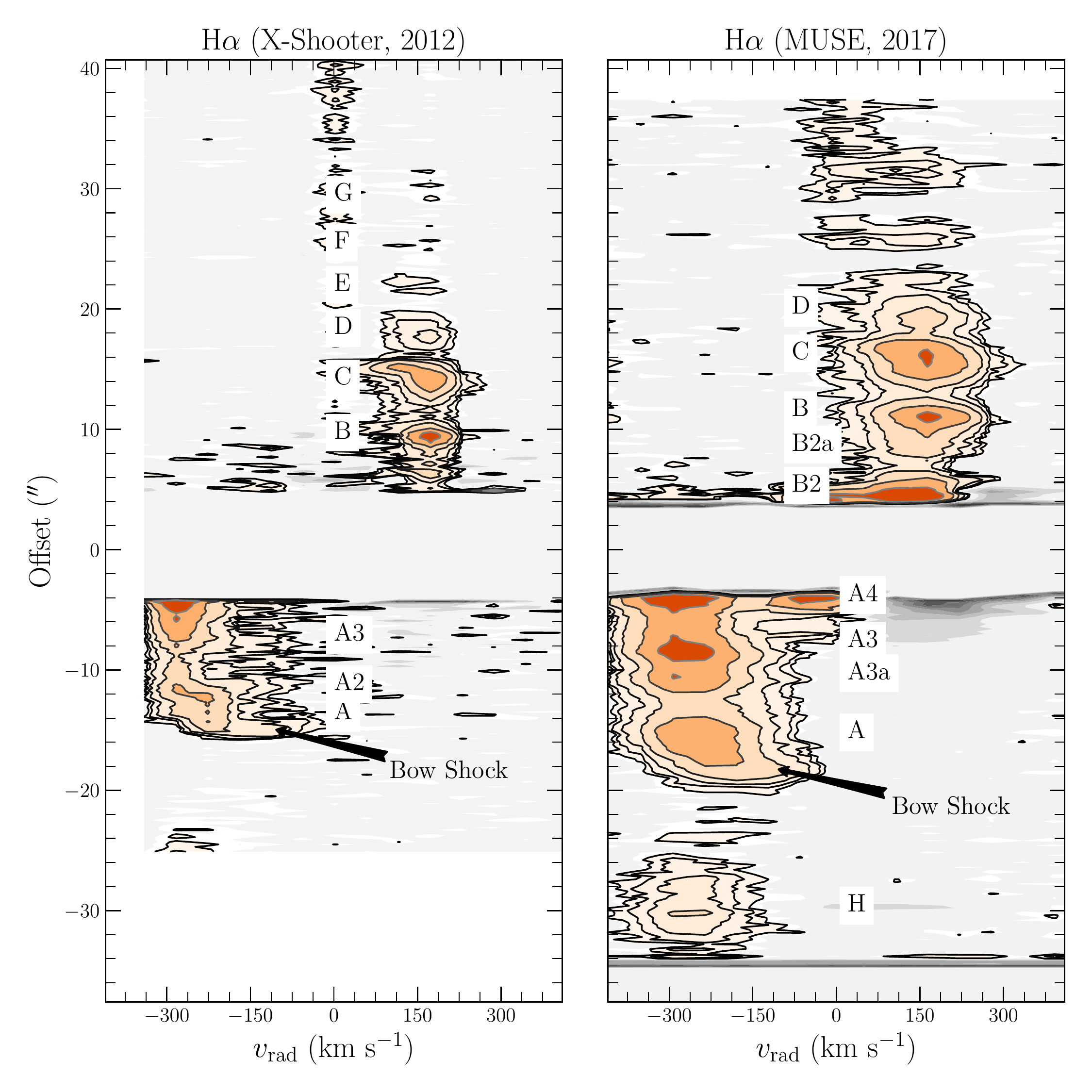}
	\caption{A comparison of the PV arrays from X-Shooter (left) and MUSE (right) for the \Ha~line. The X-Shooter spectra have been degraded to match the resolution of MUSE. The contours are on a log scale corresponding to $\log F_{\lambda} = [-17.4, -17.2, -16.9, -16.6, -16.4, -16.0, -15.7]$ for the X-Shooter data and $\log F_{\lambda} = [-18.8, -18.6, -18.3, -18.0, -17.7, -17.4, -17.0]$ for the MUSE data.}%
	\label{fig:x_comp}
\end{figure*}

Local continuum removal was performed separately for each emission line of interest, following an adaptation of the method in~\citet{AgraAmboage2009AA} which we describe in detail in Appendix~\ref{sec:appendixA}. The subtraction routine was performed with the \texttt{mpdaf}\footnote{https://github.com/musevlt/mpdaf} package in Python, producing subcubes for each desired emission line spanning smaller spectral regimes. To estimate the source centre, we have used the cubes with the shorter integration time mentioned above (with $t_{exp}=10\,$s) where saturation is minimal. A Moffat profile\footnote{For ground-based seeing-limited applications, the Moffat profile is generally more accurate for modelling the PSF; see~\citet{trujillo2001}.} was fitted to the source at each spectral step using the \texttt{moffat\_fit} routine in \texttt{mpdaf} to examine variation in the measured centre (in sky coordinates) with respect to wavelength, and the mean value in each emission region of interest was used as the source position in the data analysis. Little variance across the wavelength of the short-exposure cube was measured and this method results in a position uncertainty of approximately 10~mas, which is calculated as the standard deviation of the fitted source positions at each wavelength.

The cube was rotated so that the jet lay along the x-axis of the FOV using an affine transformation routine to ensure proper remapping of coordinates and spatial sampling. Several angles were tested and it was found that an angle of $\sim47.3 \pm 0.5^{\circ}$ (see~Section \ref{subsec:open_angle}) north through east in the plane of the sky appeared to yield the best results. This value implies a disk PA of $\sim132.7^{\circ}$, which is in agreement with the PA ($=132.2^{\circ}$) calculated by~\citet{Rich2019ApJ}. Line-fitting routines were performed with the \texttt{lmfit}\footnote{https://lmfit.github.io/lmfit-py/} package. Integrated spectro-images were extracted from the high velocity regions of the emission lines. Position velocity diagrams (PVDs) were generated by simulating a `slit' of 1\arcsec\, (5 pixels) across the approximate centre of the source. Gaussian fitting was used to estimate the knot radial velocities, knot positions, jet FWHM, and jet axis position. The results of these analyses are further discussed in Section 3. The uncertainties in the centroid positions were estimated from Eq.~\ref{eq:uncert}. The corresponding equation for the FWHM~\citep[see][Eq. A.1]{Porter2004} yields uncertainties of the order of 2~mas, and so we have instead utilised the standard error in the FWHM fits as it is more conservative.

\begin{equation}
  \centering
    \sigma_\mathrm{centroid} = \frac{\mathrm{FWHM}}{2 \sqrt{2\ln{2}} \mathrm{S/N}},
  \label{eq:uncert}
\end{equation}

\begin{table*}[ht!]
\centering
    \begin{tabular*}{2\columnwidth}{c @{\extracolsep{\fill}} cccccc}
        \hline
	\hline
        Knots & $x_t$   &  $x_t$ &  $x_t$ &Proper motion &V$_{rad}$ & V$_{rad}$   \\
           & {\small MUSE 2017 ($\arcsec$)}   & {\small XSh 2013 ($\arcsec$)} & {\small XSh 2012 ($\arcsec$)} &{\small ($\arcsec$~yr$^{-1}$)} & {\small MUSE 2017 (km~s$^{-1}$)} &{\small XSh 2012 (km~s$^{-1}$)} \\
           {\it Blue Arm} & & & &\\
        \hline
        A &  15.67 $\pm$ 0.05  &- &13.85 $\pm$ 0.03  &0.35 $\pm$ 0.01  &    255 $\pm$ 20 &260 $\pm$ 20\\
        A2 & - &- &12.12 $\pm$ 0.02  &- &- &-\\
        A3a & 10.91 $\pm$ 0.05 &-  &9.18 $\pm$ 0.13  &0.33 $\pm$ 0.01 &    270 $\pm$ 17 &-\\
        A3 & 8.34 $\pm$ 0.05   &- &6.53 $\pm$ 0.02  & 0.35 $\pm$ 0.01 & 280 $\pm$ 11 &260 $\pm$ 20\\
        A4 &  4.61 $\pm$ 0.05 &- &- &- &    280 $\pm$ 30 &- \\
         {\it Red Arm} & & &\\
        \hline
        B2  &   4.77 $\pm$ 0.05   &3.2 $\pm$ 0.03 &-   & 0.37 $\pm$ 0.02 &    125 $\pm$ 20 &-\\
        B2a &   8.83 $\pm$ 0.87  &- &6.4 $\pm$ 0.14  & 0.46 $\pm$ 0.05 &   150 $\pm$ 18 &-\\
        B &   11.11 $\pm$ 0.13 & 9.6 $\pm$ 0.03  & 9.18 $\pm$ 0.03  & 0.36 $\pm$ 0.01 &    160 $\pm$ 15 &170 $\pm$ 15\\
        C &   15.80 $\pm$ 0.23 &- &13.89 $\pm$ 0.06  & 0.36 $\pm$ 0.01 &    160 $\pm$ 15 & 160 $\pm$ 15\\
        D &   19.68 $\pm$ 0.68 &- &17.9 $\pm$ 0.64  & 0.34 $\pm$ 0.01 &    145 $\pm$ 25  & 145 $\pm$ 25\\
        \hline
	\hline
    \end{tabular*}
  \caption{Proper motion measurements calculated from the MUSE data and the two X-Shooter epochs. The radial velocities and knot offsets here are calculated from the co-added measurements of the \Ha\ and \SI\ lines. We do not include some knots previously reported due the inability to resolve their features in the MUSE data.}
    \label{knot_table}
\end{table*}

\begin{table}[h!]
    \centering
    \caption{Proper motion estimates \& jet inclination}
    \begin{tabular}{ccccc}
    \hline 
    \hline
    Knots & $v_t$   & $v_t$  & $v_{space}$ & $i_{inc}$ \\ 
      & ($\arcsec$ yr$^{-1}$) & (km s$^{-1}$) & (km s$^{-1}$) & ($^{\circ}$)\\
    \hline
    A     &  0.35 $\pm$ 0.01  & 167 $\pm$ 5 & 305 $\pm$ 10  & 57 $\pm$ 2 \\ 
    A3a   &  0.33 $\pm$ 0.01  & 160 $\pm$ 5 & 315 $\pm$ 10  & 60 $\pm$ 2 \\ 
    A3    &  0.35 $\pm$ 0.01  & 165 $\pm$ 6 & 325 $\pm$ 10  & 60 $\pm$ 2 \\ 
    B2    &  0.37 $\pm$ 0.02  & 180 $\pm$ 10 & 220 $\pm$ 20  & 35 $\pm$ 3 \\ 
    B2a   &  0.46 $\pm$ 0.05  & 220 $\pm$ 25 & 270 $\pm$ 20  & 35 $\pm$ 4 \\ 
    B     &  0.36 $\pm$ 0.01  & 175 $\pm$ 5 & 235 $\pm$ 10 & 43 $\pm$ 2 \\ 
    C     &  0.36 $\pm$ 0.01  & 175 $\pm$ 5 & 235 $\pm$ 10  & 42 $\pm$ 2 \\
    D     &  0.34 $\pm$ 0.01  & 165 $\pm$ 6 & 220 $\pm$ 20  & 42 $\pm$ 4 \\
    \hline
    \hline
    \end{tabular}
    \label{tab:proper_motions}
\end{table}


\section{Results}
\label{sec:results}

The known full extent of the HD~163296 bipolar jet was covered by the MUSE FOV, and Figure~\ref{spectro_image} presents spectro-images in \Ha~and \SI~with the brightest knots labelled. The same labelling for the knots as used by \citet{ellerbroek2014} is adopted here. The blue-shifted jet (south-west jet) boasts at least three well-defined knots within the first 20$\arcsec$ ($\sim$2000 AU), with bow-shocks at $\sim$15\arcsec\, (HH 409 A) and 30\arcsec\, (HH 409 H) from the star. HH 409 H is detected at a low signal-to-noise. The red-shifted jet displays four well-defined knots with a bow shock at a position of $\sim$ 18\arcsec\, (HH 409 C). The red-shifted knots E, F, G are detected at a low signal-to-noise. We note the lack of emission within $\sim$ 4\arcsec\ from the star due to the saturation effects discussed in Section~\ref{sec:reduction}. While the jet is observed in a significant number of emission lines~\citep[$>$20 lines in numerous species, also see][]{ChenXie2021}, we primarily focus our attention on the~\Ox, \Ha, [\ion{N}{ii}]$\lambda$6583, and \SII\ emission lines. As described, the goal of this study was to investigate the morphology and proper motions of the bipolar jets and to probe the jet axis evolution. For all the analysis discussed in this section the rotated cube was used. 

\subsection{Morphological changes and proper motions}
\label{subsec:morph}

Position velocity (PV) diagrams of the jets in \Ox, \Ha, \SI\, and \CaII\, are presented in Figure~\ref{fig:pvs1}. The radial velocities of the knots were measured from these PV diagrams by Gaussian fitting of the line emission profile at each knot position. For the \Ha~and \SI~lines, spatial profiles were extracted along the jet axis and the knot positions estimated from fits to these spatial profiles. These fits are presented in Figure~\ref{fig:full_line_fits}, and the radial velocity measurements and knot positions are given in Table~\ref{knot_table}. We make a direct comparison to the 2012 and 2013 X-Shooter data presented by \citet{ellerbroek2014}. For the blue jet knot H is not considered as it was not in the FOV of the X-Shooter observation. For the red jet only the knots within the first 20\arcsec\ are considered as knots E, F, and G are too faint in the MUSE observation. Figure \ref{fig:x_comp} compares the X-Shooter \Ha\, PV diagram of the jet with the MUSE equivalent. The reduced X-Shooter data was made available to this study by L. Ellerbroek. This comparison between the PV diagrams suggests some changes to the jet between the two observations. Knots A and A2 cannot be separated here and three new knots not previously reported are identified and labelled B2a, A3a and A4.  

To investigate these differences between the X-Shooter and MUSE data further, a spatial profile was extracted from the X-Shooter data (in the same way as was done for the MUSE data) and the knot positions fitted. The results for the \Ha\, and \SI\, lines are shown in Figure~\ref{fig:full_line_fits}. Using these fits and the position velocity diagrams, emission associated with B2a and A3a is identified in the X-Shooter data but it is not as prominent as in the MUSE data. A4 would not have been within the range of the X-Shooter observation as it lies within 5\arcsec\, of the driving source. \citet{ellerbroek2014} acquired the X-Shooter spectra by off-setting the slit with respect to the source. As a result, the inner jet region ($<$ 5\arcsec) was not covered in their observations. The previously reported knots A and A2 are not resolved in the MUSE data. This may indicate a collision in the knots, with knot A slowing post-shock and A2 colliding into it.  The knot profiles in Figure~\ref{fig:full_line_fits} demonstrate this blending and collision. 

The proper motions of the knots between the MUSE and X-Shooter observations were measured and are included in Table~\ref{knot_table}. Note that X-Shooter data taken in 2012 which covered knots B, B2 and B3, and which is reported by~\citet{ellerbroek2014}, is included in Table~\ref{knot_table}. \citeauthor{ellerbroek2014} estimated average proper motions for the red and blue arms of $v_{t, \rm red} = 0.28\pm0.01 \arcsec$ yr$^{-1}$ and $v_{t, \rm blue} = 0.49\pm0.01 \arcsec$ yr$^{-1}$ respectively. Our comparison gives average values of  $v_{t, \rm red} = 0.38\pm0.04\arcsec$ yr$^{-1}$ and $v_{t, \rm blue} = 0.34\pm0.01\arcsec$ yr$^{-1}$. We derive an average jet inclination angle $i_{\rm jet} = \arctan{ \langle v_{\rm rad}\rangle / \langle v_{\rm tan} \rangle}$ of $47\pm10^{\circ}$, which is in agreement with the findings of~\citet{Rich2019ApJ}. The uncertainties on the average proper motion and inclination values are the standard deviation of the values reported in Tables~\ref{knot_table} and~\ref{tab:proper_motions}. The changes to the knot morphologies and the significance of the new knot detections are discussed in Section~\ref{sec:propermotions}, along with the difference between the proper motions reported here and the values given by~\citet{ellerbroek2014} . 

\begin{figure*}
	\centering
\includegraphics[width=18cm, trim={3cm 1cm 3cm 0cm}]{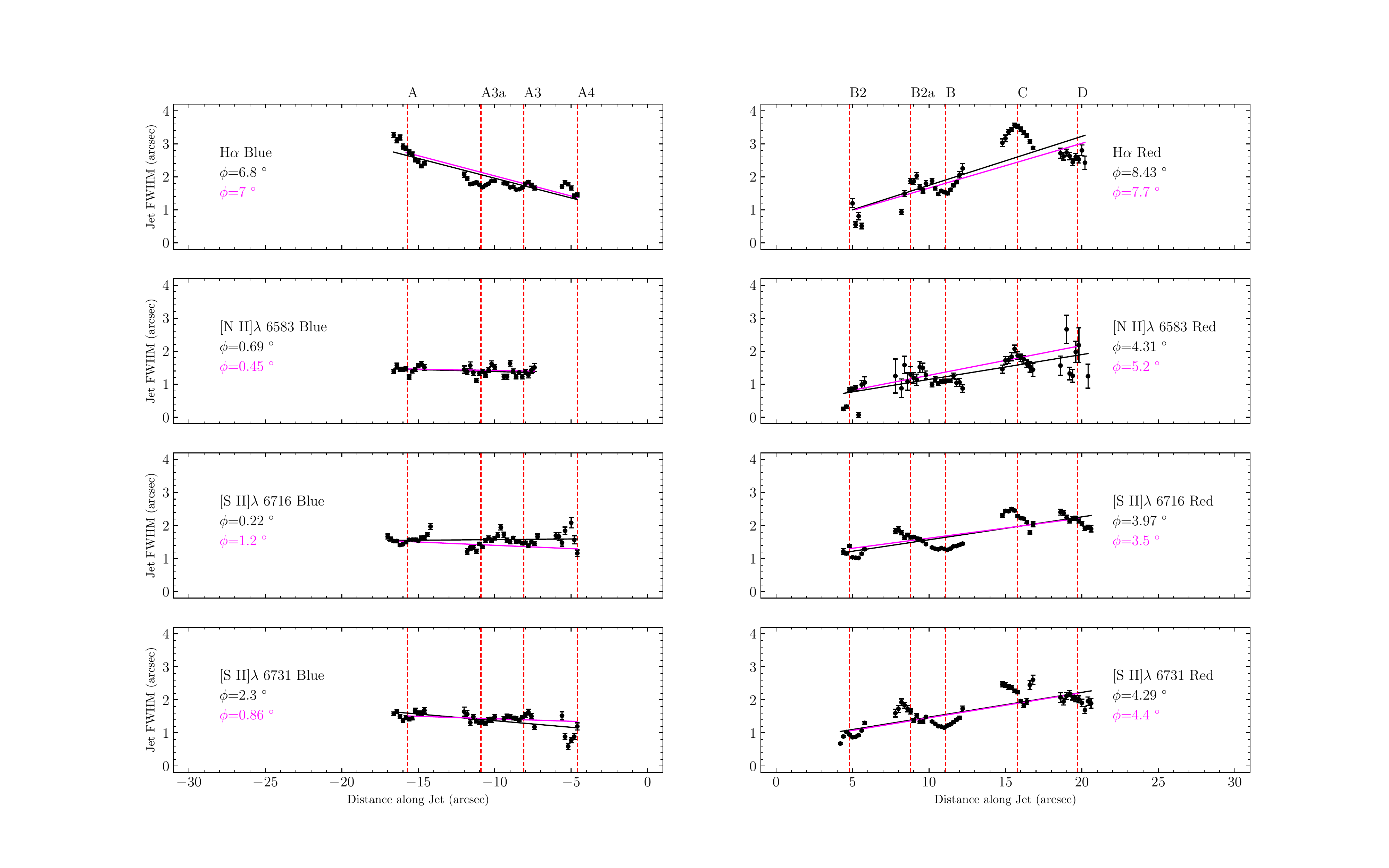} 
\caption{Intrinsic jet width and opening angle measurements for both red- and blue-shifted lobes of the outflow. The magenta lines correspond to the linear fit applied only to the knot peak positions, while the black lines correspond to a linear regression to all the data points.}%
	\label{fig:corrected_opening_angle}
\end{figure*}

\begin{figure*}
    \centering
    \includegraphics[width=\linewidth, trim={11cm 1cm 11cm 0cm}, clip=true]{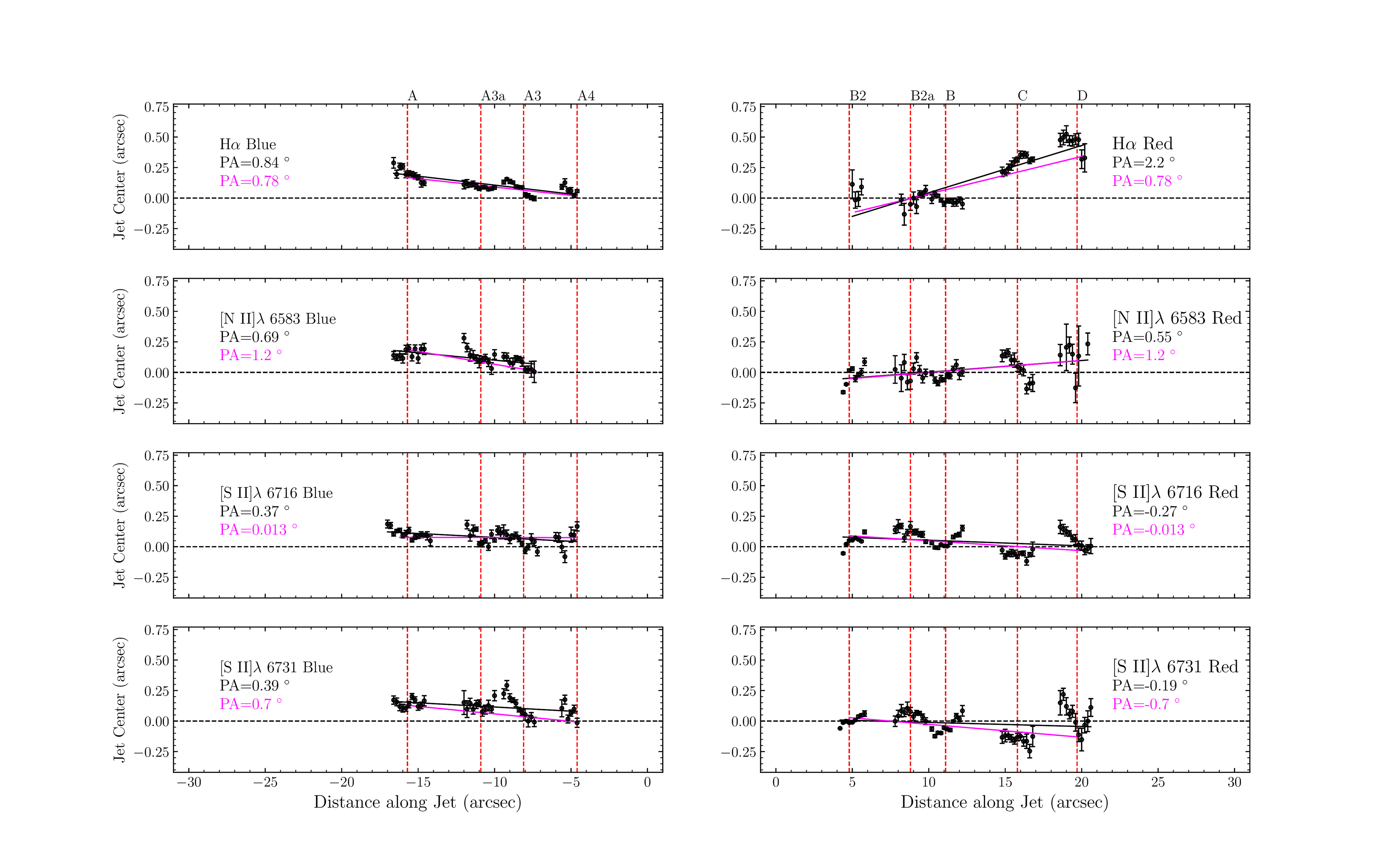}
    \caption{Gaussian centroids fitted along the jet axis for the aforementioned emission lines. As with Figure~\ref{fig:corrected_opening_angle}, the magenta lines correspond to the fit of the knot peak positions, while the black lines correspond to a fit to all the data points.}
    \label{fig:full_centroids}
\end{figure*}

\begin{figure*}
	\centering
	\includegraphics[width=18cm, trim={3cm 0cm 3cm 0cm}, clip=true]{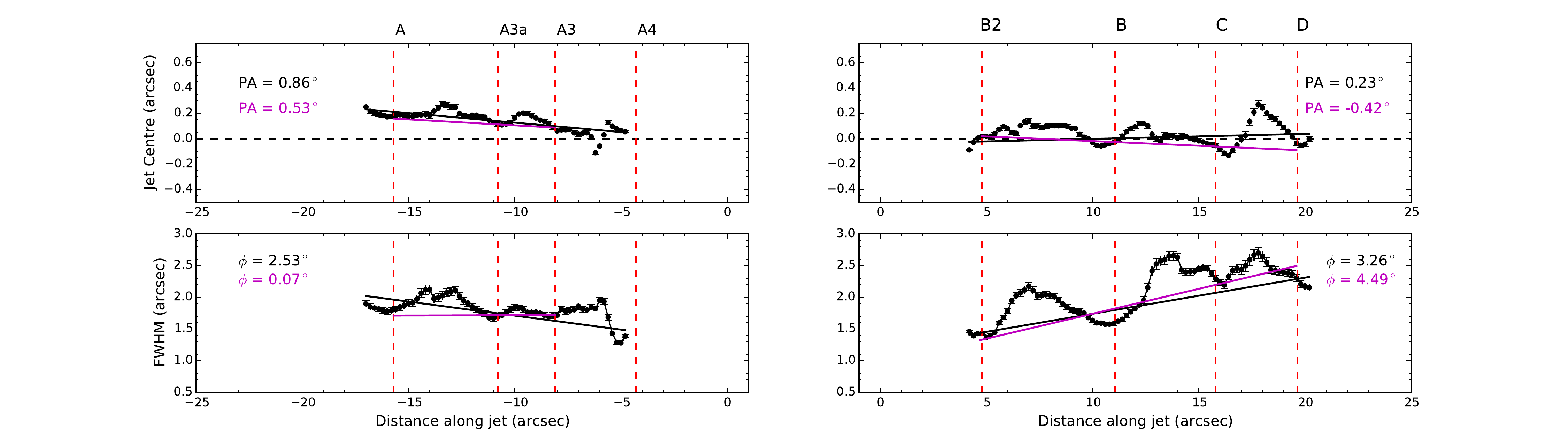}
	\caption{Top: average centroid positions of the three FELS in Fig.~\ref{fig:full_centroids}. Bottom: average FWHM measurements of the same three FELs. The data is smoothed with a 3-pixel running average.}%
	\label{fig:avg_comp}
\end{figure*}

\subsection{Jet opening angle and axis evolution}
\label{subsec:open_angle}

The analysis then turned to investigating the jet width and jet axis position with distance from the star. This analysis was done using Gaussian fitting to spatial profiles extracted perpendicular to the jet axis at each pixel along the jet. This method for measuring the jet width follows the 'jet fitting' method described in~\citet{Raga1991}. We report the intrinsic jet width taken as the FWHM with the seeing subtracted in quadrature. An average seeing of 1\farcs2 was computed by fitting Gaussian profiles to ten stars in the field and extracting their widths. As the jet is most collimated at the highest velocities, we have used spectro-images containing the high-velocity components (HVCs) of the jet, spanning $-$200~\km\, to $-$300~\km\, for the blue lobe, and 150~\km\, to 250~\km\, for the red lobe. Results for the \Ha, \NII, and \SII\, lines are presented in Figure~\ref{fig:corrected_opening_angle}. The jet is brightest in these four lines hence they are the focus here. The uncertainty at each pixel was estimated according to Equation~\ref{eq:uncert}, and after inspection of the fit at each pixel in each HVC image, we rejected all fits in which the S/N was too low ($<5$). Consequently, some parts of the inter-knot regions are excluded from Figure~\ref{fig:corrected_opening_angle}. Overall it is found that the jet width increases with distance from the star. 

Opening angles were estimated from the slopes of linear fits to the data, assuming a non-zero jet width at the source. \citet{Raga1991} found in their analysis of the knotty HH~34 jet that for every intensity maximum there is a jet width minimum. This gives the trace of the jet width with distance from the star a bumpy appearance. The authors conclude that these fluctuations are not real but are artefacts due to the knotty jet being observed under non-Gaussian PSF conditions. They recommend limiting any measurements of the jet widths to the intensity maxima, that is the knot peaks. Similar fluctuations are also seen for the HD~163296 jets and the red jet in particular. Hence, in Figure~\ref{fig:corrected_opening_angle} we show two fits. One incorporates all the data points (black line) and the second only the widths measured at the intensity peaks (magenta line). As can be seen from Figure~\ref{fig:corrected_opening_angle}, the difference between the angles estimated from the two fits is not significant. The red lobe is found to have a greater opening angle than the blue ($\sim5^{\circ}$ compared to $\sim2.5^{\circ}$ respectively), further confirming the asymmetric nature of the jet. These results are not unreasonable, as similar angles were previously estimated by~\citet{wassell2006} and such angles are generally expected as the jet is collimated within the first $\sim$100~au ($\sim$1\arcsec\ for this object) au of the source~\citep{frank2014, eisloffel2000}. 

The jet axis position as a function of distance from the source was also measured, again for the \Ha, \NII, and \SII~lines, and again using the same velocity intervals. The position of the jet axis was taken as the centroid of the Gaussian fit \citep{Murphy2021}. The results are shown in Figure~\ref{fig:full_centroids}. While the~\Ha\, results are different from the FELs, similar patterns of deviation are seen in the three FELs for both jet lobes.

Because the extended wings of the bow-shocks are typically brighter in~\Ha\, than in the FELs, we might expect that the variation of the emission centroid with distance from the star in \Ha\, would be somewhat different from the FELs.

We also note the difference in the PA of the jet lobes. Again the PA was measured by fitting all the points and then only the positions at the knot peaks.  As the PA has the same magnitude for the two jet lobes but is in opposite directions, we argue that the PA difference is a product of a slight under-rotation of $\sim 0.5^{\circ}$ of the original images rather than an intrinsic PA asymmetry. 

The agreement in the pattern of the centroid positions with distance for the FELs could lead us to conclude that we are detecting jet axis wiggling. To probe this further, we directly compare in Figure~\ref{fig:avg_comp} the Gaussian centroid measurements against the jet width measurements. For this figure, the FEL spectro-images in \NII\, \SII\, were averaged and the same process as for Figures~\ref{fig:corrected_opening_angle} and~\ref{fig:full_centroids} was followed. A running average of three pixels along the jet was used when extracting the spatial profiles, and hence the full inter-knot regions could be included. The comparison again reveals a change in the centroid of the jet emission, but we also observe a correlation between the centroid deviation, FWHM, and knot position. Notably, larger displacements and FWHM measurements are seen in the inter-knot regions. This correlation points to a different origin for the pattern we observe in the forbidden emission centroids. We further explore these alternative origins in Section~\ref{subsec:asymm}.

\section{Discussion}
\label{sec:disconc}

Similar to many other young stars, the detection of rings in the disk of HD~163296 first uncovered the possibility that its disk could host planets~\citep{grady2000, Isella2018, Zhang2018}. HD~163296 is an interesting case as the presence of more than one planet has frequently been proposed meaning that with HD~163296 we can talk of a planetary system being present. \citet{teague2018} studied the rotation curves of CO isotopologue emission relative to the Keplerian rotation. Deviations in the rotation curves were postulated to be a result of gaps carved in the gas surface density by Jupiter-mass planets. Comparison with hydrodynamic simulations suggested two Jupiter-mass planets orbiting at radii of 83~au and 137~au. \citet{pinte2018} used CO channel maps to study the Keplerian velocity of the HD~163296 disk. They find an asymmetry between the southeast and north-west sides of the disk which they argue matches a localised deviation from Keplerian velocity. Modelling shows that this could be caused by a $\approx$ 2~M$_{Jup}$ planet orbiting at a distance of 260~au. Similarly, \citet{Guidi2018} present KECK/NIRC2 L$^{\prime}$ band imaging of HD~163296 with the aim of seeking planetary mass companions and identified a point-like source at a deprojected distance of $\sim$ 67~au. Planetary isochrones suggest that the emission could be explained by the intrinsic luminosity of a 6-7~M$_{Jup}$ planet. However, the work of~\citet{Rich2019ApJ} argues against the companion proposed by~\citet{Guidi2018} while placing much more stringent mass limits ($<9$ M$_{Jup}$) on the planets proposed by~\citet{teague2018}. More recently, \citet{Rodenkirch2021} modelled the origin of the crescent shape asymmetry detected by~\citet{Huang2018} within the inner gap of the HD~163296 disk. They find that their models show that its origin could be a Jupiter-mass planet orbiting at a radial distance of about 48~au.  

All of the studies described above examined the planet hosting disk to look for evidence of a planetary system. \citet{ellerbroek2014} focused on the HD~163296 jet and discussed whether the 16~year period in the emission of knots in the jet could point to the presence of a companion. They note that the periodicity of 16~yr corresponds to a Keplerian orbit at 6~au and a radial velocity signal of a few~\km. This is the first suggestion of a possible planetary companion orbiting close to the star. Below we discuss what this MUSE study tells us about the HD~163296 planetary system. 

\subsection{The jet periodicity as inferred from the morphology}
\label{sec:propermotions}

The average proper motions measured here are 0.34~$\pm$ 0.01\arcsec\ yr$^{-1}$ and 0.38 $\pm$ 0.04\arcsec\ yr$^{-1}$ for the blue and red lobes, respectively. This is different from the values of 0.49 $\pm$ 0.01\arcsec~yr$^{-1}$ and 0.28 $\pm$ 0.01\arcsec~yr$^{-1}$ reported by~\citet{ellerbroek2014}. To calculate these values~\citet{ellerbroek2014} combined several observations of the knots, starting with HST data taken in 1998. They perform a global fit to data for each knot to derive their proper motion estimates and the results were presented in their Figure 2a. Analysing the slopes for the different knots in this figure reveals that the inner knots in the red lobe do have a proper motion larger than 0.28 $\pm$ 0.01\arcsec~yr$^{-1}$ reported from the global fit in their study, and consistent with the results presented here. The calculation of the proper motion for the blue lobe is difficult in both studies. In~\citet{ellerbroek2014} only four knots are used in the global fit to the blue lobe compared to the seven knots used in the red. Additionally, the dataset for the blue knots is less rich than for the red knots. In this study the blending of knots A and A2 complicates the estimate of the proper motions for both knots. Furthermore, while A3a and A3 are well detected in the MUSE data they are not well detected in the X-Shooter data, again contributing to the uncertainty in the proper motion estimate here. 

The observation of the knot A4 in our MUSE data raises questions about the assumed 16~yr periodicity of jet launching events proposed in~\citet{ellerbroek2014}, as the presence of this knot appears to violate this periodicity. This periodicity predicts that a new launching event should have been observed in 2018. While not observing any new knots in their 2018 HST data, \citet{Rich2020ApJ} note a $\sim32$\% chance of a launch occurring prior to their observations, but also note that they did not observe any near-infrared excess between 2016-2018, which was proposed as a signature for jet launching events~\cite{ellerbroek2014}. 

The argument against the proposed periodicity is strengthened by~\citet{ChenXie2021}, who report the detection of a knot B3 at $\sim$2\farcs5 in their MUSE narrow-field mode (NFM) observations. Under the assumption of a 16~yr periodicity, this knot should be located much closer to the source at the time of their observations ($<$1\arcsec). If this time frame applies to all the knots in the jet then a separation of $\sim$ 7\farcs8 and 4\farcs5 between each knot would be expected for the blue and red lobes respectively. The separation between knots A3 and A4 is $<$4\arcsec. However, as noted in~\citet{ChenXie2021} it could be that for B2 and A4 the full knot is not yet observed and so the separation between them and the previous knot is underestimated. If we accept A3a and B2a as real knot detections, this reduces the separation between the inner knots of both jet lobes. 

A possible alternative explanation for this is that of disk obscuration. \citet{ChenXie2021} noted that the presence of knot B3 in the NFM data is close to the edge of the disk, and suggest that if we assume the previous periodicity then this knot B3 could be interpreted as being a part of the knot B2. However, our WFM observations are unable to see this region, and~\citeauthor{ChenXie2021} further caution that instrumental issues resulted in contamination in the~\Ha\ line and they were thus unable to characterise the extinction. Additionally, near-infrared excess was observed in 2011 and 2012~\cite{ellerbroek2014, Rich2020ApJ}, strengthening the claim that B3 is a positive knot detection.

This interpretation of results appears to challenge the underlying assumptions of~\citeauthor{ellerbroek2014} of periodic, simultaneously ejected, and uniformly propagating knots. The results instead present a picture of a complicated, complex jet system that defies generalisation. Furthermore, results do not support the idea of a single companion being responsible for the emission of the jet knots.

\subsection{Origins of the apparent wiggling}
\label{subsec:asymm}

In Section~\ref{subsec:open_angle} we observed a correlation between the centroid of the jet emission perpendicular to the jet axis, the FWHM of this emission, and the knot and inter-knot regions. This correlation suggests that the origin of this pattern is not jet axis wiggling. We explore this further here.

One possibility is an intrinsic bias due to an asymmetry in the PSF. An argument for this follows from the idea that if the PSF is asymmetric, then rotating the images may introduce a bias in the spread of the data which may be falsely interpreted as a jet axis wiggling. To examine this we generated two continuum images from the rotated and non-rotated cubes on either side of the \Ha\, line in regions where jet emission was absent. We then extracted multiple field stars at the same positions in the blue and red continuum images and computed 2D Moffat fits to these sources. If the observed deviation is due to an asymmetric PSF, then a comparison of the PSF in the rotated and non-rotated cubes should reveal a discrepancy in the shapes of the profiles. In each source we observe a nearly 1:1 ratio between the FWHM$_x$ and FWHM$_y$ values (see Figure~\ref{fig:psf_fwhm_ratio}), and thus it is unlikely that we are seeing the result of some rotation artefact on an asymmetric PSF. 

We then questioned if there may be any processes which could potentially affect the photocentres of the knots. In Section~\ref{sec:propermotions} we observe a standard deviation in the calculated $i_{inc}$ for each knot of $\sim$10$^{\circ}$, and along with shock processes in the knots this may shift the emission peaks away from the jet axis. Evidence of these off-centre shocks are most clearly seen in H$\alpha$, and less-clearly in the \SII\, lines (see Figure~\ref{spectro_image}). However, our proper motions study focused primarily on comparisons with X-Shooter data, which provides us only with the magnitude of the motion and does not account for directionality beyond a general motion away from the source. Accounting for the proper motion vectors would require multi-epoch spatial imaging of the knots. Additionally, if we wished to explore the vectors and their relationship to the inclination angles of the knots, we would require each epoch of 2D images to be coupled with simultaneous spectral observations. While HST images do exist, the large gaps in time between observations would add a substantial degree of uncertainty to our calculations. While it may be that analysis of the proper motion vectors could shed more light on the origins of the observed change in the centroids with distance along the jet, such a study is beyond the scope of this paper. Instead we focus on the question of emission asymmetry in the knots themselves.

To determine whether the centroid pattern we see in Figures~\ref{fig:full_centroids} and~\ref{fig:avg_comp} is due to an intrinsic emission asymmetry in the knots, we constructed an effective PSF from the extracted field stars discussed above. Next we examined the spatial profiles transverse to each measured knot peak, focusing on knots A, A3, B, and D as these have the largest S/N. Then we extracted spatial profiles from inter-knot regions on either side of each knot, choosing the profiles again with the strongest S/N. Finally, we extract a spatial profile across the effective PSF in the same direction as the transverse knot cuts, subtract this profile from the knot and inter-knot cuts, and analyse the residuals. For this analysis we normalise all of the profiles before subtracting them, as our interest here is in the relative size and direction of the residual peaks with respect to the centre of the PSF. 

What we see from this analysis is that the residuals of the knot profiles exhibit strongly shifted peaks away from the jet axis, and that these peaks favour shifts in the negative direction with the exception of the \Ha\, line, as this line primarily traces the bow shocks in the jet. In the cases of the knots, the residuals are primarily single-peaked, and where they are double-peaked it is found that one peak is often significantly larger than the other. The residuals of the PSF-subtracted inter-knot profiles generally exhibit shifted peaks as well, but these are far more noisy and in many cases peak opposite the knot residuals. As these profiles are generally wider than the knot profiles, particularly in the \SII\, lines, we often find double-peaked residuals where one peak only slightly dominates the other. The most representative knot and inter-knot residuals are shown in Appendix~\ref{sec:appendixB}. This behaviour exists throughout inter-knot regions with no clear systematic preference towards positively or negatively shifted peaks. We additionally observe a general trend towards more extended and asymmetric wings in the inter-knot regions, with the asymmetry of the profile generally favouring positive offsets from the jet axis. 

\begin{figure*}[h]
  \centering
\includegraphics[width=\linewidth, trim={0cm 0cm 0cm 2cm}, clip=true]{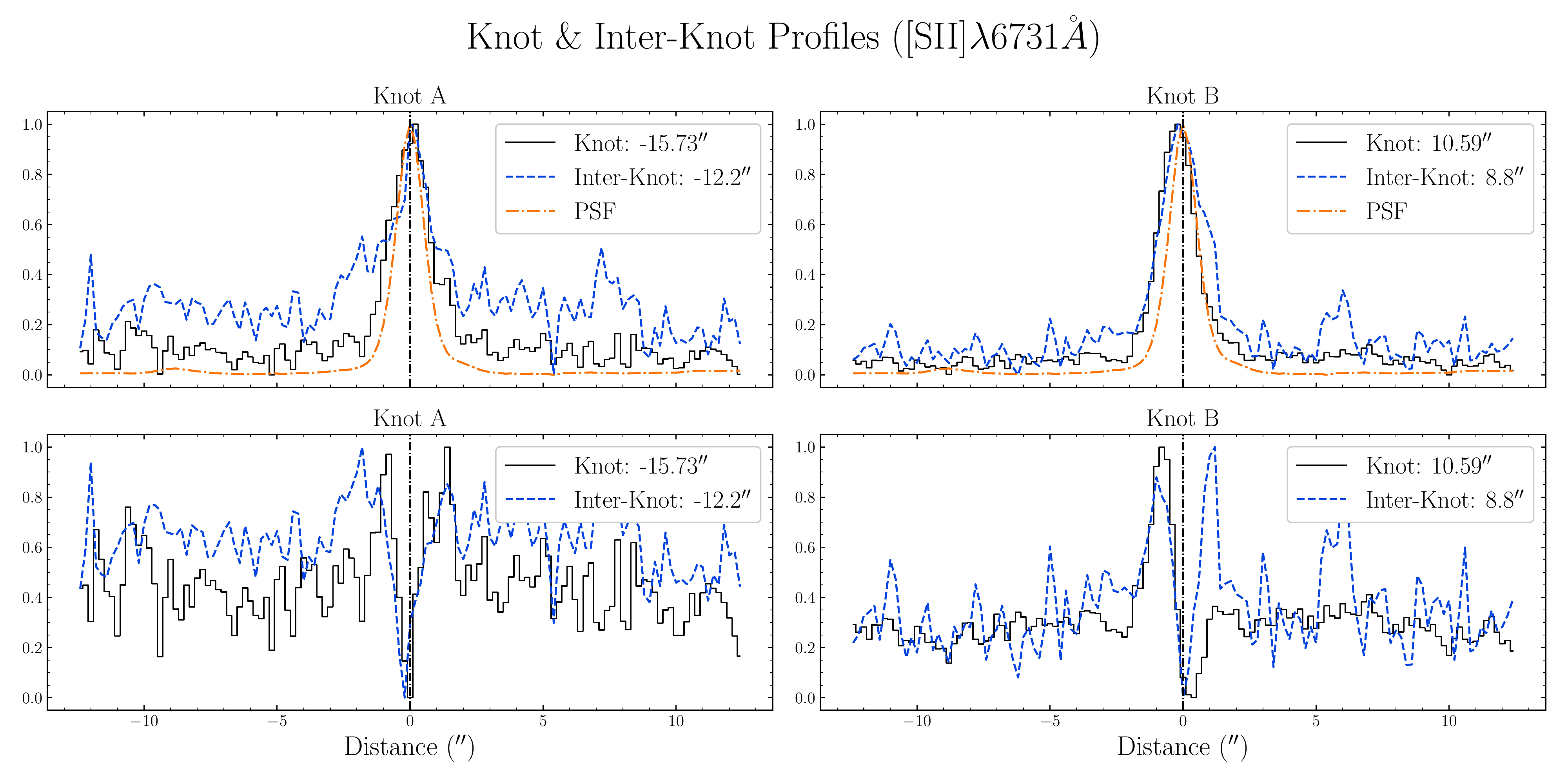}
  \caption{Spatial profiles (top) and PSF-subtracted spatial profile residuals (bottom) for the [\ion{S}{ii}]$\lambda6731$ line. Transverse cuts each with a width of 1$^{\prime\prime}$ (5 pixels) are taken across the PSF and knot/inter-knot regions specified on the plots and the profiles normalised before subtraction. We take the centre (in arcseconds) to be the location of the peak of the PSF, as we are concerned primarily with the relative shifts in knot centroid positions. Note the differences in the knots B and C compared to A and A3, suggesting that the red-shifted jet is less symmetric than its blue-shifted counterpart.}%
  \label{fig:knot_psf_residuals}
\end{figure*}

Our analysis leads us to conclude that the knots possess an intrinsic emission asymmetry that makes them present more brightly on one side of the axis than the other. This has been observed for other jets \citep{Hartigan2019} and would cause changes in the Gaussian centroids between the knot and inter-knot regions, as seen in Figure~\ref{fig:avg_comp}. Curiously, we find that the red-shifted knots have a greater tendency towards asymmetry than those in the blue. This is seen in Figure~\ref{fig:knot_psf_residuals} where the spatial profiles and PSF-subtracted residuals are compared for Knots A and B. Perhaps most interesting is that Knot A in the FELs is less pronounced in this asymmetry, while in \Ha\, the asymmetry is quite stark. It could perhaps be that the age of the knot has an impact here, such as a decrease in brightness~\citep{Rich2020ApJ}, or a more uniform distribution of the knot through time evolution. In the absence of a more detailed proper motion analysis, we conclude that an intrinsic emission asymmetry in the jet is believed to be the most reasonable explanation for our observations. An in-depth study of the shocks in the jet is planned for a future paper, which may shed further light on this phenomenon.

\subsection{The jet axis as a window on the HD~163296 system}
\label{sec:planets}

The Gaussian centroid results presented in Section~\ref{subsec:open_angle} and the conclusions drawn in Section~\ref{subsec:asymm} raise the question of whether the non-detection of a jet axis wiggle could reveal new information about the HD~163296 system. To examine this, we attempt to set bounds on the parameter space within which an orbiting companion should cause an observable wiggling, either via orbital motion or precession.

In the orbital motion scenario, the orbital velocity of the jet source, induced by a close companion, modulates the ejection velocity direction of the jet which leads to large scale jet axis wiggling. In the case of a binary orbital  plane perpendicular to the jet propagation axis, the equation of the jet axis in the plane of the sky is given by

\begin{equation}
   \centering
      y = \kappa x \cos\left( \kappa \frac{x}{r_o} -\psi \right) - r_o \sin\left(\kappa \frac{x}{r_o} - \psi \right) ,
   \label{eq:orbital_eq}
\end{equation}

\noindent where $\kappa=v_o/v_j$ is the ratio of the orbital velocity of the jet source and jet velocities, $r_o$ the orbital radius of the jet source, and $\psi$ being a phase angle~\citep{masciadri2002}.\footnote{Note here that all orbital components refer to the motion of the jet source, not the companion.} 

In the precession, a misaligned companion, either inside~\citep{ZhuMNRAS2019} or outside~\citep{terquem1999} the circumprimary disk, induces a precession of the inner disk and hence of the jet launching axis. This can be described with the expression 

\begin{equation}
   \centering
      \label{eq:precession_eq}
	 y = x \tan\beta \cos \left( \nu \left[ t - \frac{x}{v_j \cos\beta}\right]\right), 
\end{equation}

\noindent where $\beta$ is the half-opening angle of the precession cone, $\nu$ the precession frequency, and $t$ a phase parameter, with $v_j$ estimated from our proper motion study, and $\beta$ and $\nu$ observable from visual inspection of the data.

When applying each of these models, we first take the binary separation \textit{a} as a function of the orbital period,  with

\begin{figure}
    \includegraphics[width=0.9\linewidth, trim={0.5cm 0.5cm 0.6cm 0cm}, clip=true]{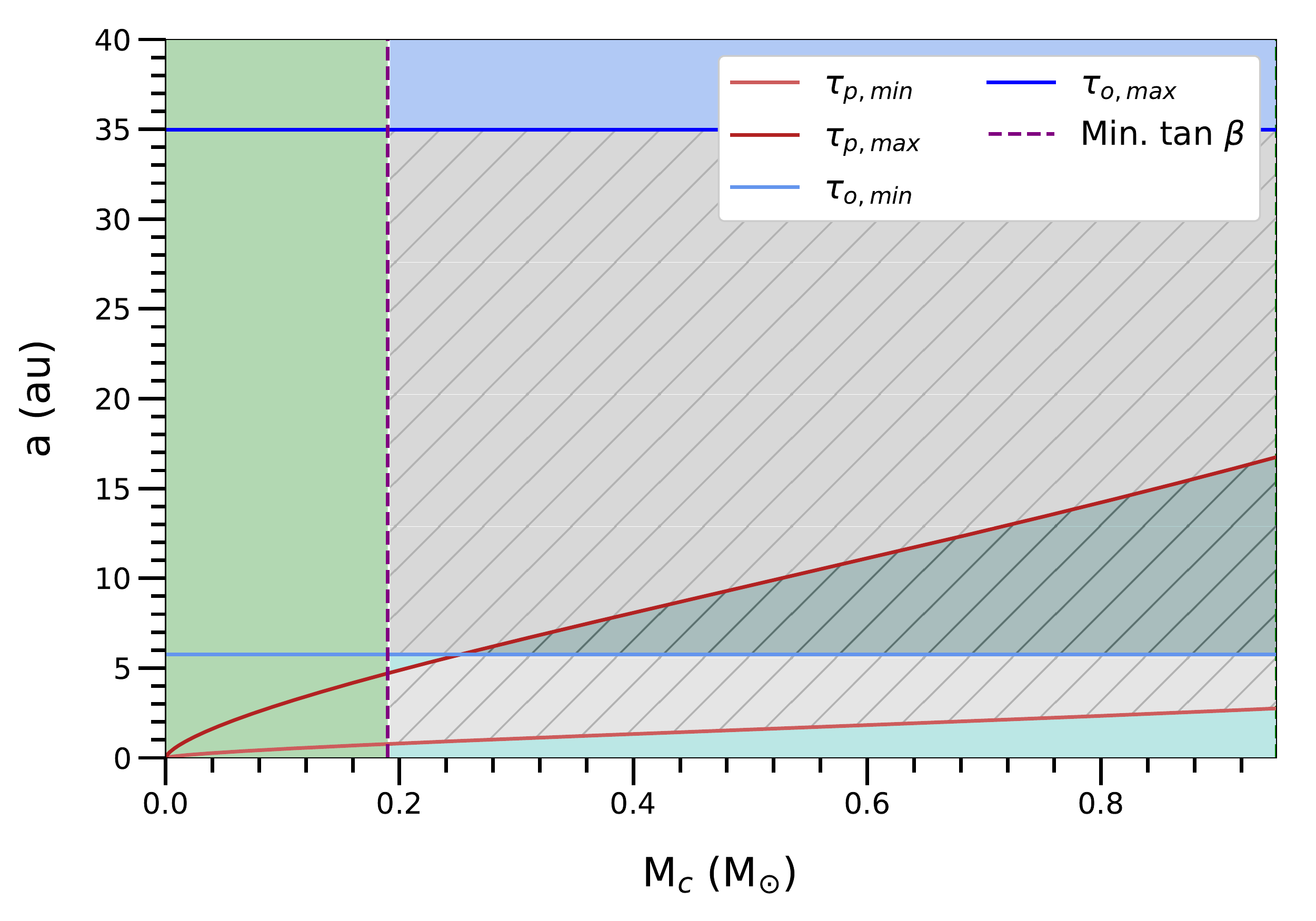}
    \caption{Parameter space of the possible companion ranges that are expected to produce a detectable jet axis wiggling in the HD~163296 jet axis. The grey hatched regions indicate parameter regions excluded by our observations (where a companion would be likely to cause a wiggling due to either orbital motion or precession, corresponding to the upper and lower hatched regions, respectively). The region shaded in green (left) contains companion objects which would produce a wiggle opening angle too small to detect in our observations; the regions shaded in blue (top) and light blue (bottom) give companions with wiggling period too long or short, respectively, to be observed in these data.}
    \label{fig:orbit_bounds}
\end{figure}

\begin{center}
\begin{equation} \label{equation:a_asmufunc}
	a = (M_{sys}\tau_{o}^{2})^{1/3},
\end{equation}
\end{center}

\noindent where $M_{sys}$ is the mass of the binary system and $\tau_o$ is the orbital period. Considering the jet proper motion and the extent of the jet which is visible in our observations, we can estimate upper and lower limits on the length scale (and hence period) it would be possible for us to detect. In the case of the orbital motion model, we can use this to set $\tau_{o}$ directly. In the alternate case, this is taken as the precession period $\tau_p$. The ratio of orbital and precession periods is itself a function of the companion mass ratio $\mu = M_{c}/M_{sys}$, and we can therefore obtain $\tau_{o}$ from

\begin{equation}
	\begin{aligned}
		\label{equation:brokendisk_period}
			\dfrac{\tau_{o}}{\tau_{p}} = \dfrac{3}{8} \dfrac{\mu}{\sqrt{1 - \mu}} \sigma^{3/2} \cos{i_{p}},
	\end{aligned}
\end{equation}

\noindent using the equation for a broken-disk scenario as described in \citet{ZhuMNRAS2019} where $\sigma=R/a$ is the ratio of the radius of the circumstellar disk to the binary separation. We note that $\sigma = 1$, and $i_{p}$ is the angle between the orbital angular momentum vector of the companion and the rotation axis of the jet, which is taken to be $<$ 1$^{\circ}$ similar to the precession angle. We choose this relationship as the short precession periods in this case would imply a small companion orbiting at separations of $\sim$1 au or less. In the case of HD~163296, we trace approximately 15\arcsec\, of the jet within our observations. Taking an approximate proper motion of 0\farcs4 yr$^{-1}$, we estimate the maximum period that would show a clear deflection of the knots to be 150 years ($\lambda_{wiggle}$ = 60\arcsec), and the minimum period to be approximately 10 years ($\lambda_{wiggle}$ = 4\arcsec). These provide the upper and lower bounds of the parameter space excluded by our observations, as illustrated in Figure~\ref{fig:orbit_bounds}.

We then consider the possibility of a wiggling in the jet on a length scale within our observable range, but with an amplitude too small to be detected. Fitting the apparent wiggle in the jet gives a half-opening angle $\beta$ of 0.2$^{\circ}$. Since any wiggle smaller than this is likely to go undetected, and since $\beta$ is related to the maximum orbital velocity of a possible companion by

\begin{equation}
	\begin{aligned}
		\label{equation:kappa_limit}
		\kappa \le \tan \beta,
	\end{aligned}
\end{equation}

\noindent where $\kappa = v_{o}/v_{j}$, we can combine this in to set an upper limit on the maximum orbital velocity of a companion too small to produce a wiggling. This gives us a lower mass limit in Figure~\ref{fig:orbit_bounds}. Finally, we can set the maximum companion mass from physical considerations, as it is unlikely for the companion to be more massive than the jet source (i.e. half the mass of the total system). We can therefore construct bounds for the companion parameters excluded by the non-detection of a wiggling in this jet.

From Figure~\ref{fig:orbit_bounds}, the bounds set by these limits exclude jet wiggling due to the presence of objects greater than about 0.1~\Msun\, with separations between 1-35~au (the regions shaded in grey). A companion below 0.1~\Msun\ would not produce a large enough wiggling amplitude to be detected, while a companion of mass >0.1~\Msun\ at smaller or larger separations would produce a wiggling either too short or too long to be observed within the jet length we trace. We note that Figure~\ref{fig:orbit_bounds} illustrates the two parameter regions which individually would be expected to produce a detectable wiggling motion, given the assumptions outlined above for each scenario. This analysis shows that we would not expect to detect any of the planets inferred to be present from studies of the HD~163296 accretion disk using the jet wiggling method. It also tells us that if the periodicity of the jet holds, and is caused by a companion at 6~au, the companion must have a mass $<$0.1~\Msun\ i.e. a brown dwarf or planet, otherwise we would have expected to detect it. 

As a follow-up question, we consider under what conditions we would expect to detect a planetary-mass companion from the HD 163296 jet. Therefore we explore the anticipated jet shape for a 1.9~\Msun\ primary with a 10~M$_{Jup}$ companion ($\mu = 0.005$). Such a low-mass object is most likely to cause a measurable wiggle via precession in a close orbit and so we focus on this scenario. Assuming a precession period of 10 years (that is, within the range observable from this jet), the precession model would imply a $v_{o}$ of 0.7 \km (with $a =$ 0.09 au, from Equations~\ref{equation:a_asmufunc}  and~\ref{equation:brokendisk_period}). With a jet velocity of 235 \km, we can then infer a minimum half-opening angle $\beta \geq$ 0.17$^{\circ}$ due to the orbital motion. The observed $\beta$ due to the precession wiggling may be significantly larger as this is determined by the precession axis. Additionally, this minimum angle is inversely proportional to the jet velocity, and so if $v_{j} \sim 100$ \km, then $\beta \geq$ 0.4$^{\circ}$.

On the other hand, if we consider observing an orbital motion wiggle, then $v_o$ = 0.1 km s$^{-1}$, and the expected half-opening angle is 0.02-0.06$^{\circ}$ (again, with larger values corresponding to lower jet velocity). Since the wiggle opening angle directly corresponds to the ratio of orbital to jet velocity, this is not merely a lower limit. If we consider a longer wiggling period of 100 years, for example, we find even smaller opening angles; conversely, if we assume a lower system mass $\sim$1~\Msun, we obtain slightly larger opening angles.

Thus, we can first infer that we are more likely to detect a small object through precession than directly through orbital motion, as for any small companion the corresponding precession wiggle will be larger in amplitude. Additionally, the wiggle will be easier to detect for short precession periods (hence very short orbital periods), low-velocity jets, and lower-mass jet sources. This poses two limitations: first, that the precession wiggle will only give us an upper limit to the companion mass and is therefore unlikely to directly constrain this to a value of a few Jupiter masses or less; second, this method is most likely to detect planetary-mass objects at separations of $<$ 1 au \citep{Erkal2021AA}.

\section{Conclusions}
\label{sec:conc}
A morphological study of the HD~163296 jet, including its width and axis position with MUSE in wide-field mode is presented here. We report the presence of three new knots we have labelled B2a, A3a, and A4. A proper motions study found that the number of knots and the separation between the knots challenges the 16 year periodicity proposed by~\citet{ellerbroek2014}. This contradiction is supported by the observations of~\citet{ChenXie2021}, and provides evidence of a complex jet structure. Examination of the jet axis along the direction of the outflow suggests that both lobes are effectively perpendicular to the circumstellar disk, but that there exists an asymmetry in the opening angle of the jet between the two lobes ($5^{\circ}$ in the blue and $2.5^{\circ}$ in the red). This examination also revealed a regular deviation of the jet emission centroid from the jet axis, which has been observed in some other jets and has been argued as a potential indication of a companion \cite{masciadri2002, Erkal2021AA, Murphy2021}. Thus we have devoted much focus to understanding the source of this pattern. We directly compare the emission centroids, the jet FWHM, and the knot positions and rule out a jet axis wiggling. We construct an effective PSF to further explore the possibility of either an instrumental effect or an intrinsic emission asymmetry in the knots. We find no significant asymmetry in the PSF, and analysis of PSF-subtracted spatial profiles along the knot and inter-knot regions supports the existence of asymmetric knot emission. We ultimately conclude that the centroid deviation is due to this asymmetry. We note, however, that in-depth multi-epoch analysis of the proper motion vectors of the knots and a detailed examination of the shock structure in the jet may shed more light on possible origins of our observed centroid deviation pattern. Overall for HD~163296, the non-detection of a jet axis wiggling allows us to rule out the presence of a companion greater than about 0.1~\Msun\ with separations between 1-35~au.

\begin{acknowledgements}
	The authors extend our thanks to Lucas Ellerbroek for providing us with reduced X-Shooter data, and for his thoughtful discussions and ideas regarding our own research. We also wish to thank Chen Xie for sharing crucial NFM mode observations with us and offering insight into our analysis. A. Kirwan wishes to acknowledge funding through the John and Pat Hume Doctoral Scholarship at Maynooth University, Ireland. P. C. Schneider gratefully acknowledges support by DLR Grant 50 OR 2102. Finally, we thank the anonymous referee for their insightful comments that greatly improved the manuscript.
\end{acknowledgements}


\bibliography{_ref}
\bibliographystyle{aa}


\clearpage
\begin{appendix}
\section{Local continuum contribution removal}

\label{sec:appendixA}

In order to obtain the clearest view of the jet emissions, it is important to accurately remove the local stellar contribution at each spatial pixel across the image. We achieve this by adapting the method described in~\citet{AgraAmboage2009AA}. We select a region close to the source (though not fully on-source, due to the saturation issues discussed in Section~\ref{sec:reduction}) that is free of jet emission, and extract a spectrum centred on the emission line of interest. The exact pixel location of this reference spectrum was chosen by visual inspection and thus varied from sub-cube to sub-cube, but was generally within 2\arcsec\, south-east of the source. A low-order polynomial is fitted to this reference spectrum to create a reference baseline. Next, we iterate over the cube spaxel by spaxel and fitted a baseline to the background spectrum at that coordinate, and calculate a scaling array by dividing this local baseline by the reference baseline. Finally, the stellar reference spectrum is multiplied by this scaling array and subtracted from the spectrum at that pixel. Strong emission and absorption features are masked when computing the baseline to prevent biasing the polynomial fit.

It is important to note that the FWHM of the PSF is known to vary such that the total signal in any given pixel is dependent on wavelength, and care must be taken to avoid biasing the subtracted spectrum such that it presents bluer or redder than it actually is. To mitigate this issue, we require that the above operation be performed very locally. In Section~\ref{sec:propermotions} we explore possible asymmetries in the PSF and find a nearly 1:1 ratio between the FWHM$_x$ and FWHM$_y$. Furthermore, examining the PSF images in Figure~\ref{fig:psf_fwhm_ratio} reveals that the FWHM varies by $\sim1\%$ between the blue and red images, indicating only a small variance over the given regime. For this reason, we have limited our local continuum removal to small line widths of $\sim$100~\AA\, to minimise bias in the spectrum. By scaling the reference baseline to each local pixel over small wavelength bins and removing its contribution from the local spectrum, we can achieve the best image of the intrinsic emission from the jet. 

Despite masking the stronger emission and absorption features when computing the scaling array, some strong negative residuals remained after the subtraction process, particularly around the H$\alpha$ line. The source spectrum at this emission features dips in the wings profile, which produce large residuals in areas close to the saturated regions due to over-subtraction. A median filter was tested to minimise the effect of these residuals, but they were ignored as their proximity to the regions of saturation results in a larger uncertainty in the analysis of the knots closest to the source. Visual inspection of histograms of the affected regions allowed us to estimate a $\sigma$-value for each sub-cube, and we chose instead to mask pixels where values fell below $-5\sigma$.

\section{Emission Asymmetries}
\label{sec:appendixB}
\begin{figure*}[t]
    \centering
    \includegraphics[width=0.8\linewidth,trim={2cm 0cm 2cm 0cm}, clip=true]{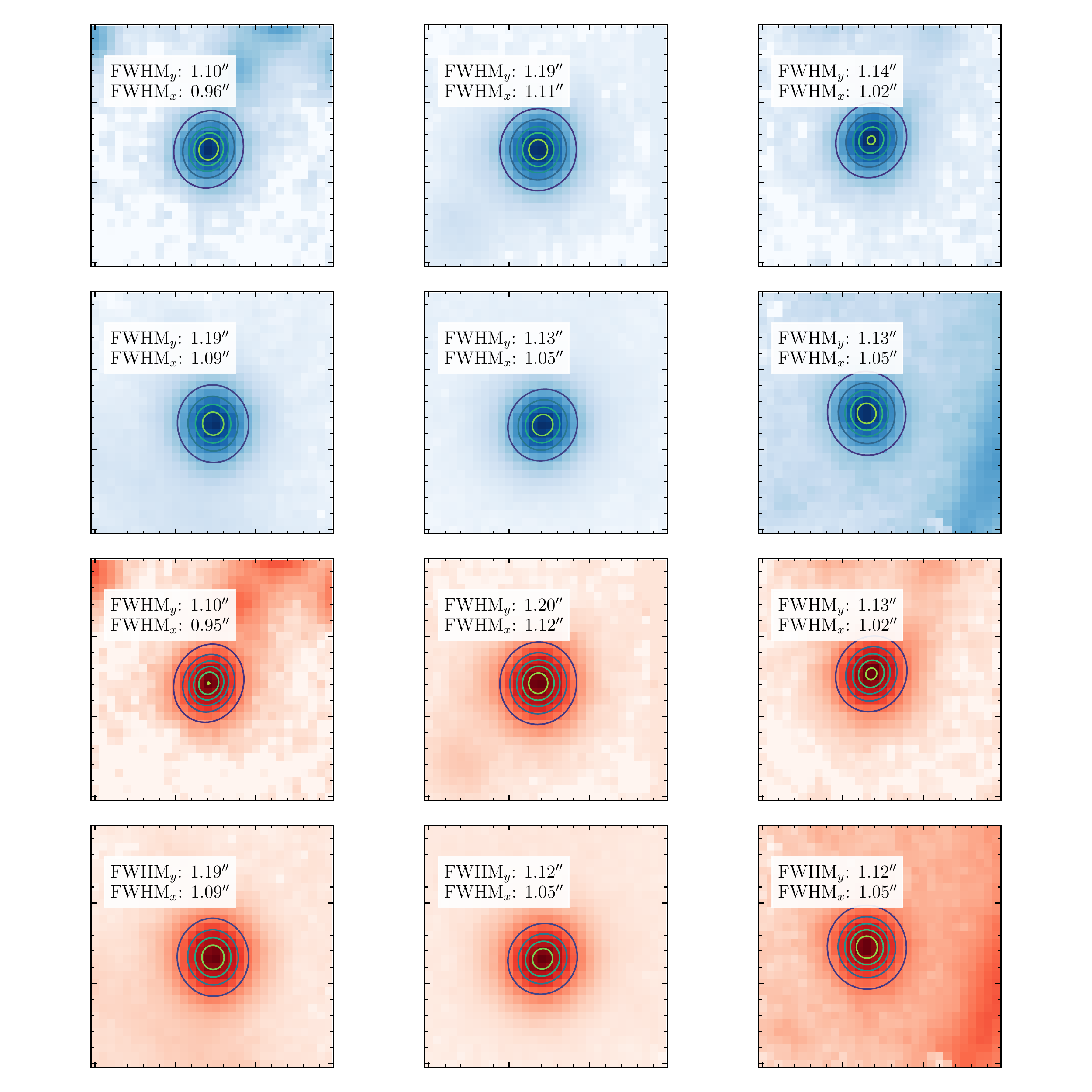}
    \caption{A sample of source stars extracted from blue continuum (top) and red continuum (bottom) images relative to the \Ha\, line. Each continuum image is integrated over a 20\AA\, bin. We fitted each source with a 2D Moffat profile to examine the FWHM ratio, and use these to build an effective PSF. With these fits, we find a ratio FWHM$_y$ $/$ FWHM$_x = 1.09\pm0.06$.}
    \label{fig:psf_fwhm_ratio}
\end{figure*}

\begin{figure*}[htpb]
  \centering
  \includegraphics[width=\linewidth, trim={0cm 0cm 0cm 2cm}, clip=true]{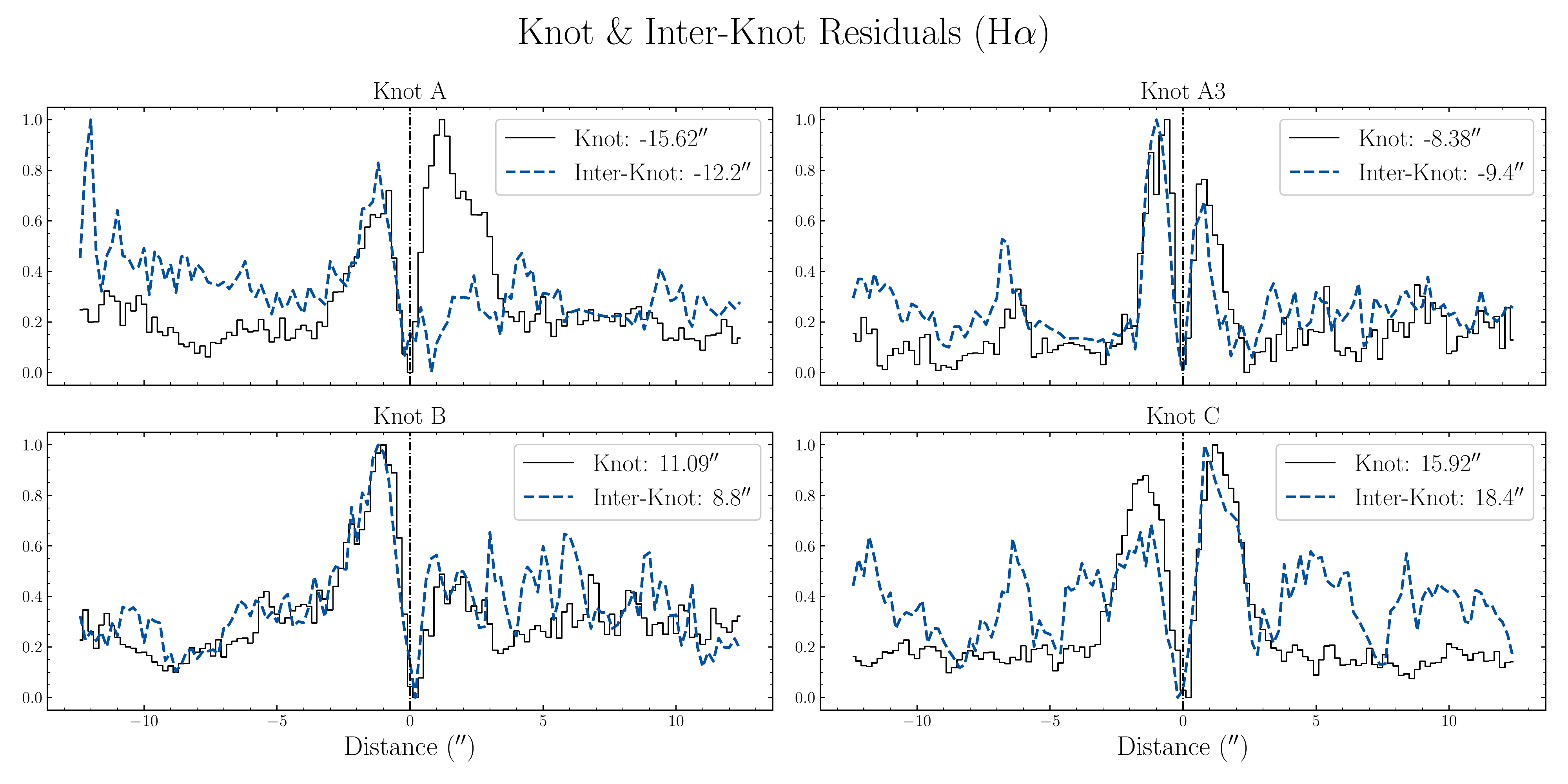}
  \caption{PSF-subtracted residuals of knot and inter-knot regions in the H$\alpha$ line. As with Figure~\ref{fig:knot_psf_residuals}, profiles are summed over a 1$^{\prime\prime}$ width and normalised before subtraction. Note that Knot A in this line is markedly different from the same knot in the FELs, and that the relative offsets in all the knots are much larger than their FEL counterparts. This is most likely due to the behaviour of the bow shocks in the knots.}%
  \label{fig:residual_Halpha}
\end{figure*}

\begin{figure*}[htpb]
  \centering
  \includegraphics[width=\linewidth, trim={0cm 0cm 0cm 2cm}, clip=true]{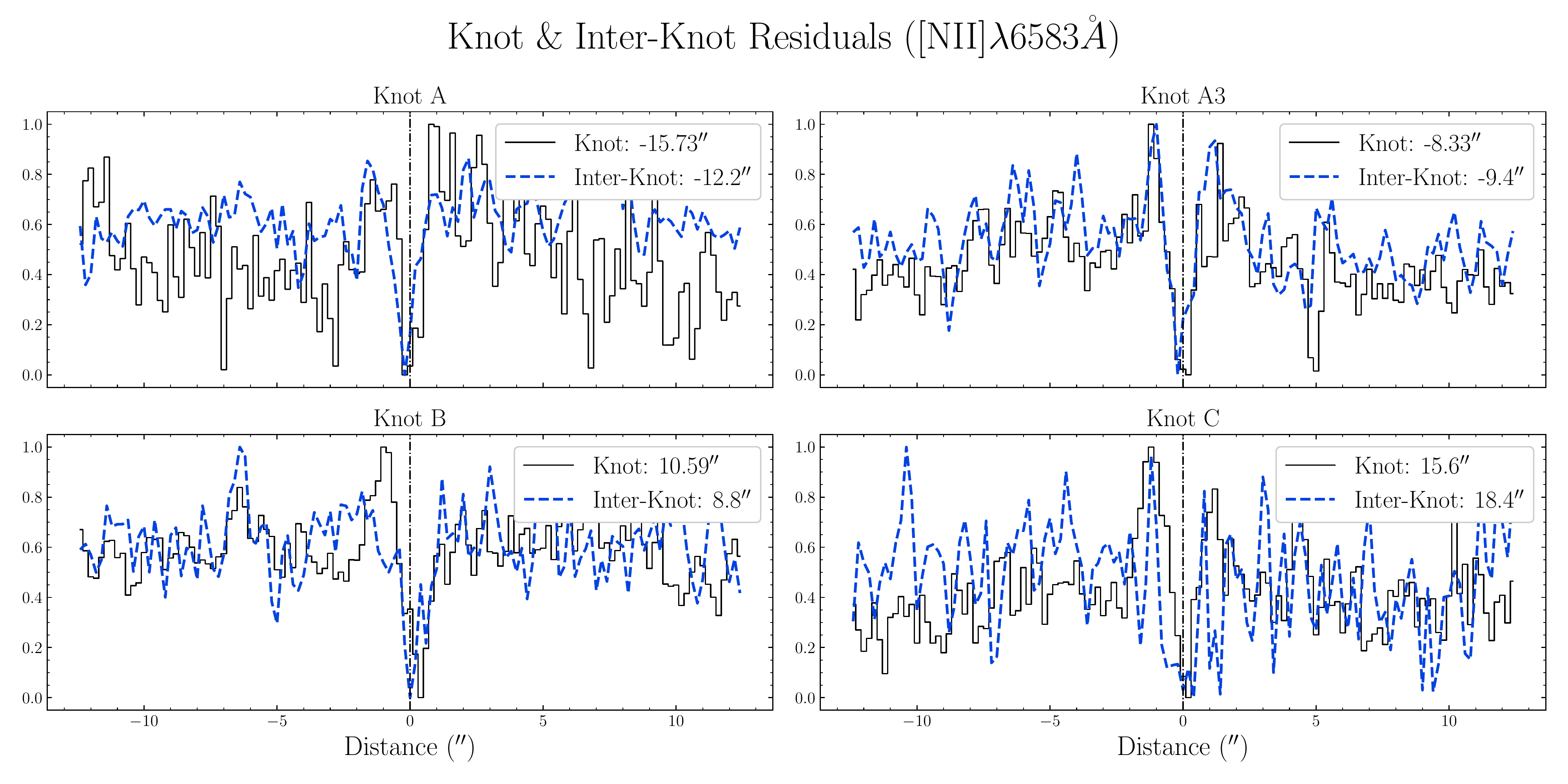}
  \caption{PSF-subtracted residuals of knot and inter-knot regions in the [\ion{N}{ii}]$\lambda$6583 line, using the same method outlined above. The [\ion{N}{ii}] line is rather faint compared to the other FELs, and so we observe less deviation from the jet axis in this emission.}%
  \label{fig:residual_NII}
\end{figure*}

\begin{figure*}[htpb]
  \centering
  \includegraphics[width=\linewidth, trim={0cm 0cm 0cm 2cm}, clip=true]{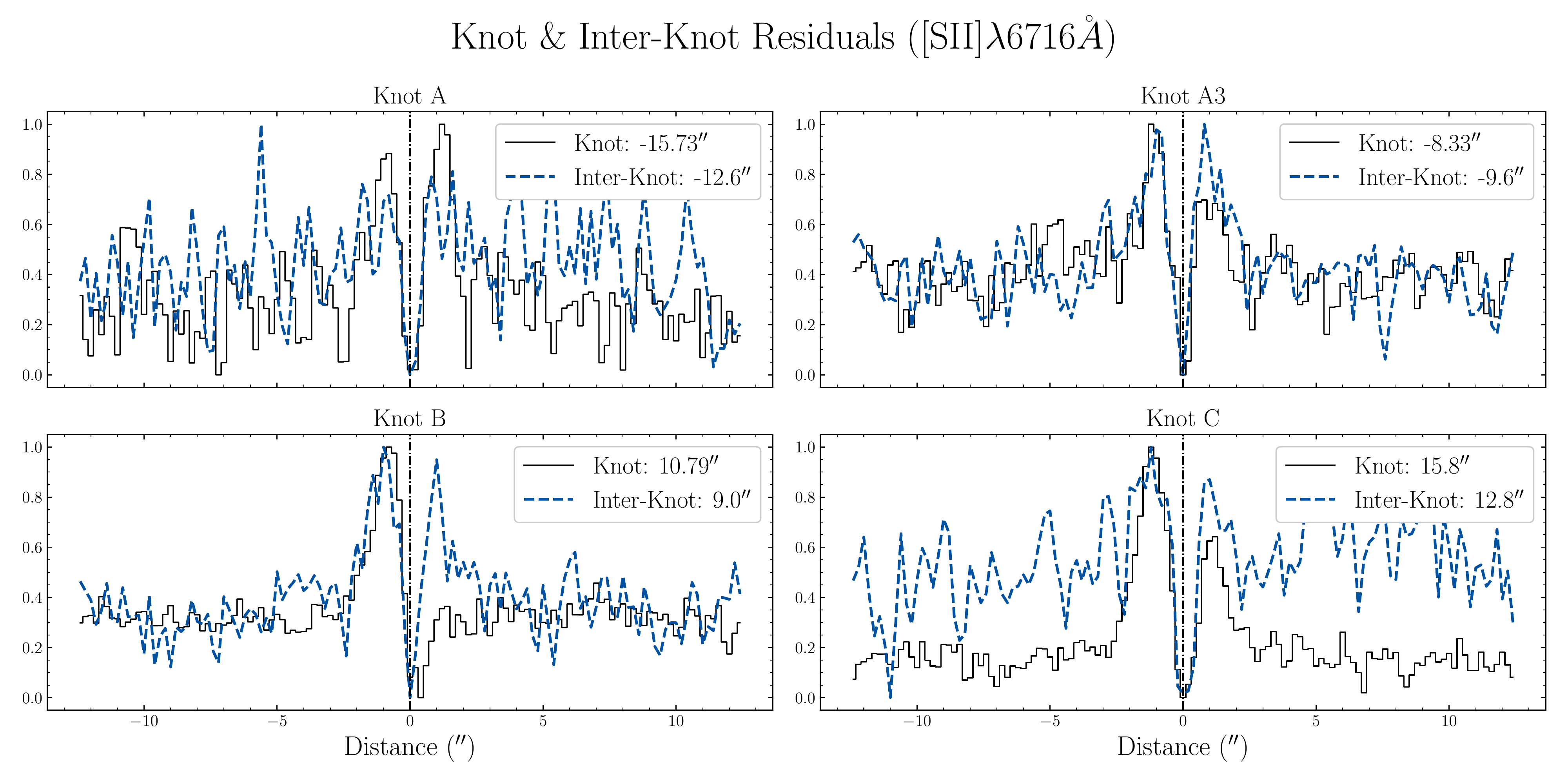}
  \caption{PSF-subtracted residuals of knot and inter-knot regions in the [\ion{S}{ii}]$\lambda$6716 line, following the same method as above. We see knots A, A3, and C each exhibiting double-peaked residuals, with negative peaks dominating A3 and C. As shown in Figure~\ref{fig:residual_Halpha}, knot A does not appear to follow this trend towards negative offsets.}%
  \label{fig:residual_SIIa}
\end{figure*}

\begin{figure*}[htpb]
  \centering
  \includegraphics[width=\linewidth, trim={0cm 0cm 0cm 2cm}, clip=true]{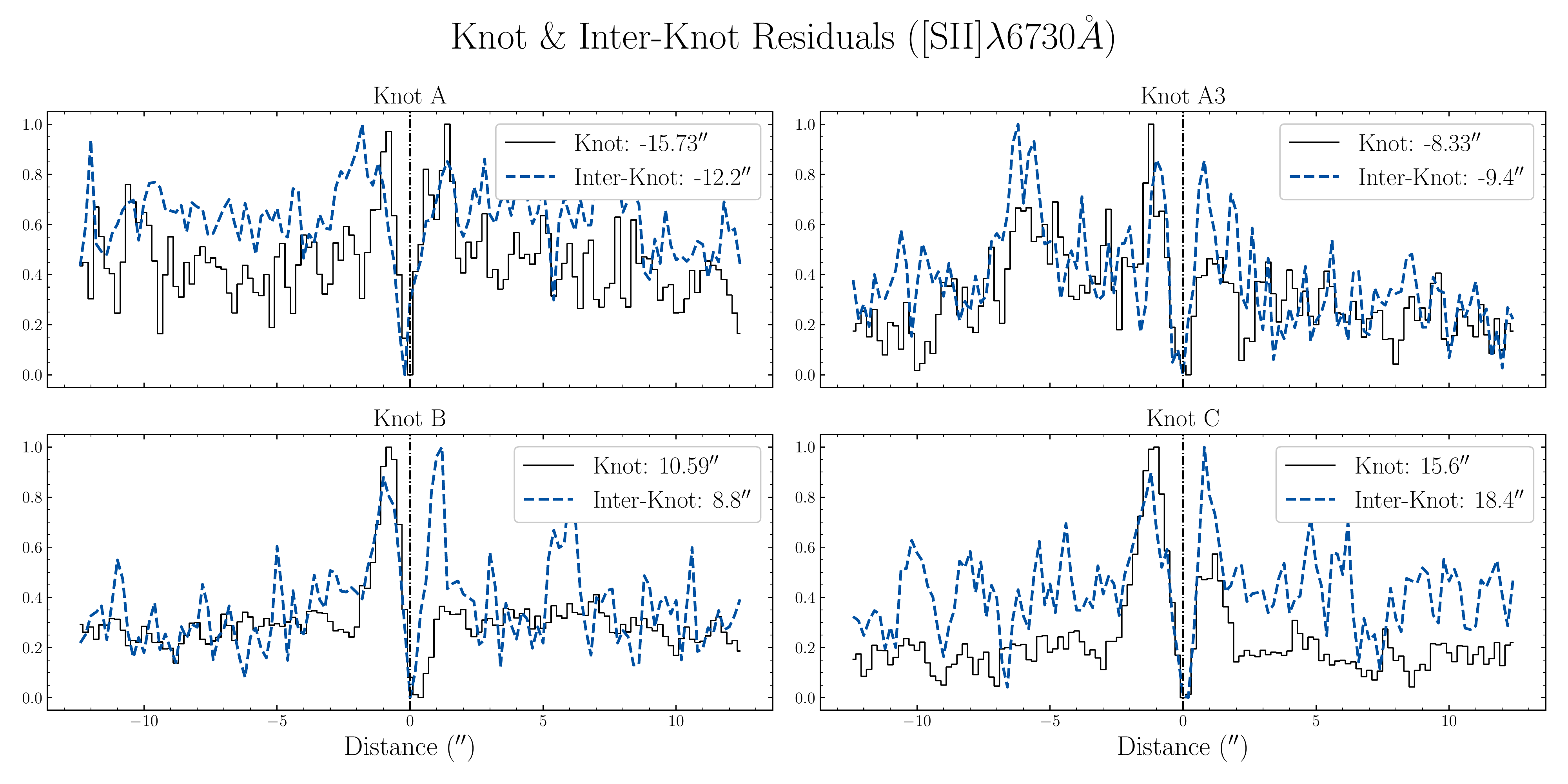}
  \caption{PSF-subtracted residuals of knot and inter-knot regions in the [\ion{S}{ii}]$\lambda$6731 line using the method discussed in~\ref{subsec:asymm}. As with the above [\ion{S}{ii}] line, we see a clear preference towards negative-shifted residuals with the exception of Knot A.}%
  \label{fig:residual_SIIb}
\end{figure*}

Figure~\ref{fig:psf_fwhm_ratio} depicts multiple field stars extracted around the H$\alpha$ line. Both blue and red images are integrated over 20~\AA\, bins, with the lower limit  of the blue emission and upper limit of the red emission spanning a total line width of 100~\AA. Analysis of this image is discussed in Section~\ref{sec:propermotions} and Appendix~\ref{sec:appendixA}, and this image highlights the importance of limiting our local continuum contribution removal routine to small line widths. Figures~\ref{fig:residual_Halpha} through~\ref{fig:residual_SIIb} feature the PSF-subtracted residuals of the spatial profiles discussed in Section~\ref{subsec:asymm}. The residuals are all normalised, so only the relative contribution of the subtracted profiles is shown. 

\end{appendix} 
\end{document}